\pdfoutput=1
\documentclass[usenatbib]{mn2e}
\usepackage{hyperref}
\usepackage{apjfonts}
\usepackage{amssymb}
\usepackage{amsmath}
\usepackage{ctable}
\usepackage{url}
\usepackage{breakurl}
\usepackage{fixltx2e} 

\newcommand{\be}{\begin{equation}}
\newcommand{\ee}{\end{equation}}

\newcommand{\paperone}{Paper {\small I}}

\newcommand{\gizmourl}{\burl{http://www.tapir.caltech.edu/~phopkins/Site/GIZMO.html}}
\newcommand{\vspacerpostplot}{\vspace{-0.4cm}}

\newcommand{\divB}{\nabla\cdot{\bf B}}

\newcommand{\divBi}{(V\,\divB)_{i}}

\newcommand{\dB}{\nabla\otimes{\bf B}}

\newcommand{\comment}[1]{}
\newcommand{\divBnorm}{h_{i}\,|\divB|_{i} / |{\bf B}|_{i}}
\newcommand{\dberr}{\divBnorm}
\newcommand{\dBerr}{\divBnorm}

\newcommand\plotonesize[2]
 {\centering \leavevmode \includegraphics[width={#2\columnwidth}]{#1}}
\newcommand{\plotsidesize}[2]
 {\centering \leavevmode \includegraphics[width={#2\textwidth}]{#1}}
\newcommand{\acknowledgments}{\begin{small}\section*{Acknowledgments}\end{small}}
\newcommand\altaffilmark[1]{$^{#1}$}
\newcommand\altaffiltext[1]{$^{#1}$}
\title[Constrained-Gradient MHD]{A Constrained-Gradient Method to Control Divergence Errors in Numerical MHD\vspace{-0.5cm}}

\author[Hopkins]{
\parbox[t]{\textwidth}{ 
Philip F. Hopkins\altaffilmark{1}\thanks{E-mail:phopkins@caltech.edu}
} 
\vspace*{6pt} \\
\altaffiltext{1}{TAPIR \&\ The Walter Burke Institute for Theoretical Physics, Mailcode 350-17, California Institute of Technology, Pasadena, CA 91125, USA\vspace{-1.1cm}} \\
}

\date{Submitted to MNRAS, July, 2015\vspace{-0.6cm}}
\begin{document}
\maketitle
\label{firstpage}

\begin{abstract}
In numerical magnetohydrodynamics (MHD), a major challenge is maintaining $\divB=0$. Constrained transport (CT) schemes achieve this but have been restricted to specific methods. For more general (meshless, moving-mesh, ALE) methods, ``divergence-cleaning'' schemes reduce the $\divB$ errors; however they can still be significant and can lead to systematic errors which converge away slowly. We propose a new constrained gradient (CG) scheme which augments these with a  projection step, and can be applied to any numerical scheme with a reconstruction. This iteratively approximates the least-squares minimizing, globally divergence-free reconstruction of the fluid. Unlike ``locally divergence free'' methods, this actually minimizes the numerically unstable $\divB$ terms, without affecting the convergence order of the method. We implement this in the mesh-free code {\small GIZMO} and compare various test problems. Compared to cleaning schemes, our CG method reduces the maximum $\divB$ errors by $\sim1-3$ orders of magnitude ($\sim2-5$\,dex below typical errors if no $\divB$ cleaning is used). By preventing large $\divB$ at discontinuities, this eliminates systematic errors at jumps. Our CG results are comparable to CT methods; for practical purposes, the $\divB$ errors are eliminated. The cost is modest, $\sim 30\%$ of the hydro algorithm, and the CG correction can be implemented in a range of numerical MHD methods. While for many problems, we find Dedner-type cleaning schemes are sufficient for good results, we identify a range of problems where using only Powell or ``8-wave'' cleaning can produce order-of-magnitude errors.
\end{abstract}

\begin{keywords}
methods: numerical --- hydrodynamics --- instabilities --- turbulence --- cosmology: theory
\vspace{-1.0cm}
\end{keywords}

\vspace{-1.1cm}
\section{Introduction}
\label{sec:intro}

Magnetohydrodynamics (MHD) is essential to many physical problems, and (because the equations are non-linear) often requires numerical simulations. But this poses unique challenges. Naive discretizations of the MHD equations lead to violations of the ``divergence constraint'' ($\divB=0$); unfortunately, certain errors related to non-zero $\divB$ are numerically unstable (they corrupt the solution even at infinite resolution). As such, many methods have been developed to control them. 

The CT method of \citet{evans:1988.constrained.transport}, and related vector-potential/flux-central difference methods can conserve an initial $\divB=0$ to machine precision each timestep; however, while it may be possible to extend these methods in principle \citep[see][]{mocz:2014.constrained.transport.mhd}, it has thus far only been practical to implement for real problems in regular, fixed-grid (Eulerian) type schemes. It is also (often) computationally expensive.

For other (e.g.\ mesh-free, moving-mesh, arbitrary Lagrangian-Eulerian, smoothed-particle) methods, ``divergence-cleaning'' schemes are popular. The \citet{powell:1999.8wave.cleaning} or ``8-wave'' cleaning simply subtracts the unstable $\divB$ terms from the equation of motion \citep[for a discussion of the instability, see][]{brackbill.barnes:projection.divb.mhd.control,toth:2000.divB.constraint,yang.li:divB.stability}; this cures the instability and restores Galilean invariance, but does not actually reduce $\divB$ (it only ``transports'' the errors). Many studies have shown that certain types of problems, treated only with this method, will converge to the wrong solution \citep{toth:2000.divB.constraint,mignone:2010.ctu.mhd.divb.constraint,mocz:2014.constrained.transport.mhd,hopkins:mhd.gizmo}; moreover the subtraction necessarily violates momentum conservation, so one would like to minimize the subtracted terms. More sophisticated cleaning schemes have been proposed; many of which follow \citet{dedner:2002.divb.cleaning.scheme} and add a scalar-field and set of source terms which transport the divergence in waves and damp it (and correct behavior in shock jumps). However, this still requires a finite ``response time'' to damp $\divB$ (so may act ``too slowly'' in discontinuities), and is reactive (dissipating, rather than preventing errors); as such it is less than ideal. Nevertheless, these schemes have been applied across a wide range of methods. 

Alternatively, projection schemes following \citet{brackbill.barnes:projection.divb.mhd.control} take the solution at each timestep, and project it onto a globally divergence-free basis (or equivalently, solve for the divergence-free component of the fluxes, and subtract off the other components).\footnote{This is generally done by representing ${\bf B}$ as ${\bf B}=\nabla \times {\bf A} + \nabla \phi$, solving for the vector/scalar fields ${\bf A}$ and $\phi$, subject to some constraints (e.g.\ $\phi$ is constrained to minimize its least-squares value/integral over volume), then taking ${\bf B}\rightarrow {\bf B}-\nabla\phi$.} This can reduce $\divB$ below a desired tolerance at each timestep, acts ``instantly,'' and as shown in \citet{toth:2000.divB.constraint} preserves the convergence order of the method. However, it is expensive (requires a global sparse matrix inversion every timestep), can become unacceptably inefficient when adaptive/hierarchical timesteps and/or non-regular mesh geometries are used, is not compatible with arbitrary slope/flux limiters, and the inversion itself can become unstable under certain circumstances. Hence the application of these methods has been limited.\footnote{Of course, many further divergence-control schemes have been proposed \citep[see e.g.][]{swegle:1995.sph.stability,monaghan:2000.sph.tensile.instability,borve:2001.sph.mhd.regularization,maron:2003.gradient.particle.mhd,price:2006.sph.mhd.neutron.star.mergers,rosswog:2007.sph.mhd,price:2008.sph.mhd.star.clusters,dolag:2009.mhd.gadget}. However, many of these examples either fail to cure the numerical tensile instability, are zeroth-order inconsistent (so do not actually converge), are so diffusive and/or expensive as to be impractical, or cannot represent non-trivial magnetic field configurations. We will therefore not consider them further.}

In this paper, therefore, we propose a hybrid constrained-gradient (CG) scheme which combines some advantages of the schemes above. We implement this in the multi-method Lagrangian MHD code {\small GIZMO},\footnote{A public version of this code is available at \gizmourl. Users are encouraged to modify and extend the capabilities of this code; the development version of the code is available upon request from the author.} and test the scheme in a wide range of problems.

\vspace{-0.5cm}
\section{Numerical Methodology}
\label{sec:methods}

\subsection{The Problem}

In most finite-volume Godunov-type methods, the MHD equations are a set of hyperbolic partial differential conservation equations for an element (particle or cell) $i$, surrounded by elements $j$, which take the discrete form 
\begin{align}
\label{eqn:final}
\frac{d}{dt}(V\,{\bf U})_{i} + \sum_{j}\,{{\bf F}}_{ij}\cdot {\bf A}_{ij} = (V\,{\bf S})_{i}
\end{align}
where $V$ is the element volume, ${\bf U}=(\rho,\,\rho\,{\bf v},\,\rho\,e,\,{\bf B},\,\rho\,\psi)$ is a vector of primitive variables (mass/momentum/energy density, magnetic field, scalar fields), ${\bf S}$ is a vector of source terms, ${\bf A}_{ij}$ is an oriented ``effective face area'' defining the interaction surface between the elements, and ${\bf F}_{ij}$ is the relevant flux, computed by solving the appropriate Riemann problem at the face.\footnote{Throughout, ``$\otimes$'' denotes the outer product, ``$\cdot$'' the inner (dot) product, and ``$:$'' the double-dot product ${\bf A}:{\bf B} = \sum_{ij} A_{ij}\,B_{ij}$. We use $\sum_{ij}$ to denote double-summation $\sum_{ij} \equiv \sum_{i}\sum_{j}$.}

The inputs to the Riemann problem at ${\bf A}_{ij}$ are the reconstructed quantities ${\bf U}_{R}$ and ${\bf U}_{L}$, which are the extrapolated values of ${\bf U}$ on the ``$i$-side'' and ``$j$-side'' of the face, respectively. In second-order methods, if the point at which the Riemann problem is solved (usually the midpoint of the face) lies at coordinates ${\bf x}_{ij}$, then ${\bf U}_{R} = {\bf U}^{\prime}_{i} + \tilde{\phi}\,\left(\nabla \otimes {\bf U} \right)_{i}  \cdot \left( {\bf x}_{ij} - {\bf x}_{i} \right)$, 
where $(\nabla \otimes {\bf U} )_{i}$ is the gradient tensor calculated at element $i$, $\tilde{\phi}$ is an appropriate slope-limiter which restricts the gradient values to prevent the creation of new maxima/minima, and ${\bf U}^{\prime}_{i}$ denotes the value of ${\bf U}_{i}$ which is appropriately time-centered for the numerical time integration scheme.

In MHD, we wish to preserve $\divB=0$. Since there are an infinite number of valid definitions of the {\em discrete} gradient operator, it is impossible for any non-trivial field configuration to satisfy this under all operators. The {\em relevant} definition of $\divB$, in most second-order finite-volume schemes, is something like
\begin{align}
\label{eqn:divb.definition}
\divBi &\equiv -\frac{1}{2}\sum_{j} \left( {\bf B}_{R} + {\bf B}_{L}  + \frac{\psi_{L}-\psi_{R}}{c_{h,\,ij}}\,\hat{\bf A}_{ij}\right)\cdot {\bf A}_{ij}
\end{align}
where the (optional) $\psi$ terms arise from divergence-cleaning schemes such as that in \citet{dedner:2002.divb.cleaning.scheme}, following a source function proportional to $\divBi$. The averaging of ${\bf B}_{R}$ and ${\bf B}_{L}$ appears because the one-dimensional Riemann problem requires that the ${\bf B}$ component normal to the face be constant. This definition represents the numerically unstable terms, which are subtracted in cleaning schemes \citep{powell:1999.8wave.cleaning,dedner:2002.divb.cleaning.scheme}; also, since this represents a surface integral, maintenance of $\divBi=0$ according to this definition is physically equivalent to magnetic flux conservation. Ideally, we would always have ${\bf B}_{R} = {\bf B}_{L}$ and $\divBi=0$ (hence also $\psi=0$). This is exactly what CT schemes try to ensure. 

Note that variations of Eq.~\ref{eqn:divb.definition} are possible, and can (if carefully defined) represent valid gradient definitions. We will use Eq.~\ref{eqn:divb.definition} as the basis of our subsequent derivations because it is common to many codes, and specifically it is the definition of relevance for the {\small GIZMO} code which we will use for our tests. However it is straightforward to modify our subsequent derivations for a modified form of Eq.~\ref{eqn:divb.definition}. 

\vspace{-0.5cm}
\subsection{Locally Divergence-Free Methods}
\label{sec:local.divb.free}

We have some freedom in the choice of discrete approximation for the ${\bf B}$-field gradients, $\dB$. For second-order methods, we must choose a definition such that the errors in the reconstruction scale $\propto h^{2}$ (where $h$ is the linear element size) in smooth flows, but this still allows considerable flexibility. 

To make progress, we will adopt the gradient estimator in {\small GIZMO}, a moving least-squares estimator. For a scalar $f$, this is 
\begin{align}
\label{eqn:gradient} (\nabla f)_{i}^{\alpha} &= \sum_{j}(f_{j} - f_{i})\,\left({\bf W}_{i}^{-1}\right)^{\alpha\beta}\,({\bf x}_{j}-{\bf x}_{i})^{\beta}\,\omega_{j}({\bf x}_{i}) \\
\label{eqn:gradient.matrix} {\bf W}_{i}^{\alpha\beta} &\equiv \sum_{j}\,({\bf x}_{j}-{\bf x}_{i})^{\alpha}\,({\bf x}_{j}-{\bf x}_{i})^{\beta}\,\omega_{j}({\bf x}_{i})
\end{align}
here we assume an Einstein summation convention over the indices $\beta$ corresponding to the spatial dimensions, and $\omega_{j}({\bf x}_{i})$ is an (arbitrary) weight function defined in \paperone. This estimator is second-order accurate for an arbitrary mesh configuration, minimizes the (weighted) least-squares deviation $\sum_{j}\omega_{j}\,| f_{i} + \nabla f_{i}\cdot ({\bf x}_{j}-{\bf x}_{i}) - f_{j}|^{2}$, and has been applied in a wide range of different numerical methods \citep[see e.g.][]{onate:1996.fpm,kuhnert:2003.finite.pointset.method,maron:2003.gradient.particle.mhd,luo:2008.compressible.flow.galerkin,lanson.vila:2008.meshfree.consistency}. 

It is straightforward to constrain this to obtain the locally divergence-free solution: i.e.\ $(\dB)_{i}$ such that $(\divB)_{i} \equiv \sum_{k} (\dB)_{i,\,kk} = 0$. In the least-squares formulation, this is just a constrained least-squares problem -- we seek the matrix $(\dB)_{i}$ constrained to have $(\divB)_{i}=0$ which minimizes the (weighted) square deviation of ${\bf B}_{i} + ({\bf x}_{j}-{\bf x}_{i}) \cdot (\dB)_{i} - {\bf B}_{j}$. 

A similar approach is to calculate the gradient projected onto a set of divergence-free basis functions. The matrix formulation above implicitly adopts the Cartesian polynomial basis functions $(1,\,x,\,y,\,z,\,x^{2},\,xy,\,xz,..)$, but a divergence-free basis can be chosen instead and used for the reconstruction. 

The problem with both of these is as follows. If we temporarily ignore the slope-limiter and $\psi$ terms in Eq.~\ref{eqn:divb.definition}, then $\divBi$ is just
\begin{align}
\label{eqn:approx.divb}
\divBi = -\frac{1}{2}& \sum_{j} {\Bigl[} {\bf B}_{i} + (\dB)_{i}\cdot ({\bf x}_{ij}-{\bf x}_{i}) \\
& + {\bf B}_{j} +  (\dB)_{j}\cdot ({\bf x}_{ij}-{\bf x}_{j}) {\Bigr]} \cdot {\bf A}_{ij} 
\nonumber + \mathcal{O}\left(\tilde{\phi},\,\psi \right) 
\end{align}
It is obvious from this expression that ensuring $(\divB)_{i}=0$ does not ensure $\divBi=0$; in fact, it does not even necessarily decrease $\divBi$.
\footnote{It is easy to verify this. Consider a trivial case: a 2D, perfectly regular lattice of equally-spaced elements so ${|\bf A}_{ij}|=A$ and $|{\bf x}_{ij}-{\bf x}_{i}|=\delta x$, and take ${\bf x}_{i}=0$. Assuming $(\divB)_{i}=0$, Eq.~\ref{eqn:approx.divb} simplifies to $\divBi =-A\,[(B_{x}(\delta x,\,0) - B_{x}(-\delta x,\,0) + B_{y}(0,\,\delta x) - B_{y}(0,\,-\delta x)]/2$. Assume $B_{y} = f(y)$, so $B_{x} = B_{x,\,0}(y) - (df/dy)\,x$; then $\divBi = A\,[(df/dy)\,\delta x + (f(\delta x)-f(-\delta x))/2] \ne 0$ for any non-linear $f$. Or simply assume a noisy, but constant-mean field, such that $(\nabla \otimes {\bf B})_{i} = 0$; then $(\divB)_{i}=0$, but $\divBi \propto \sum_{j}({\bf B}_{i} + {\bf B}_{j})\cdot {\bf A}_{ij}$, which can only vanish for special configurations of ${\bf B}_{j}$ and ${\bf A}_{ij}$.}

\vspace{-0.5cm}
\subsection{An Approximate, Globally Divergence-Free Method}

Still, these locally divergence-free methods suggest a solution. Assume, for now, that we are estimating the gradient $(\divB)_{i}$ of element $i$, and all other gradients in the system have been determined; also for now neglect the slope-limiter and $\psi$ terms so $\divBi$ follows Eq.~\ref{eqn:approx.divb}. Then we see that we can, in fact, ensure $\divBi=0$, provided that $(\dB)_{i}$ satisfies: 
\begin{align}
\label{eqn:constraint}
(\dB)_{i} : &\sum_{j} ({\bf x}_{ij} - {\bf x}_{i})  \otimes {\bf A}_{ij} = \\ 
\nonumber &- \sum_{j} \left[ {\bf B}_{i} + {\bf B}_{j} + (\dB)_{j} \cdot ({\bf x}_{ij} - {\bf x}_{j}) \right] \cdot {\bf A}_{ij}
\end{align}
Or, in component form, 
\begin{align}
\sum_{ab}\,&(\dB)^{ab}_{i}\,Q^{ab} = \sum_{ab}\,(\partial B^{a} / \partial x^{b})_{i}\,Q^{ab} = S_{0} \\ 
{\bf Q} &\equiv \sum_{j} ({\bf x}_{ij} - {\bf x}_{i})  \otimes {\bf A}_{ij} \\ 
\label{eqn:S0} S_{0} &\equiv - \sum_{j} \left[ {\bf B}_{i} + {\bf B}_{j} + (\dB)_{j} \cdot ({\bf x}_{ij} - {\bf x}_{j}) \right] \cdot {\bf A}_{ij}
\end{align}
This is a single scalar equation: so it only constrains one degree of freedom of $(\dB)_{i}$.

Say that our ``preferred'' gradient, in the absence of this constraint, is $(\dB)_{i,\,0}$. This is the gradient (accurate to the desired level of reconstruction) calculated by whatever default method, before any consideration of the constraint. 
Then define 
\begin{align}
(\dB)_{i} &= (\dB)_{i,\,0} + {\bf {G}} \
\end{align}
where ${\bf G}$ is a correction term that ensures $(\dB)_{i}$ satisfies Eq.~\ref{eqn:constraint}. We would like to make $(\dB)_{i}$ as close to $(\dB)_{i,\,0}$ as possible, so we choose the tensor ${\bf G}$ which satisfies Eq.~\ref{eqn:constraint} while minimizing some ``penalty function,'' which in this paper we take to be the least-squares deviation 
\begin{align}
f_{\rm penalty} &\equiv {\bigl |}(\dB)_{i} - (\dB)_{i,\,0} {\bigr |}^{2} = |{\bf {G}} |^{2} = \sum_{ab}\,\left(G^{ab} \right)^{2}
\end{align}
This gives the solution
\begin{align}
{\bf G} &= -{\bf Q}\,\left(\frac{S_{0} + (\dB)_{i,\,0}:{\bf Q}}{{\bf Q}:{\bf Q}} \right) \\ 
G^{ab} &= -Q^{ab}\,\left(\frac{S_{0} + \sum_{cd}\,(\dB)_{i,\,0}^{cd}\,Q^{cd}}{\sum_{cd}\,(Q^{cd})^{2}}\right)
\end{align}

We note that while this changes the gradient from our ``preferred'' estimator, and will, of course, change the errors in the solution, the term ${\bf G}$ is explicitly minimized, and is sourced by non-zero $\divB$ errors. In any well-designed numerical scheme, these errors are of the convergence order of the code -- therefore, although this alters the gradients, it is easy to show that it does {\em not} change the convergence order. 

There are two problems that prevent this from exactly providing $\divBi=0$. ({\bf 1}) First, ${\bf G}$, and hence each $(\dB)_{i}$, depends on the neighboring gradients $(\dB_{j})$, through Eq.~\ref{eqn:S0}. ({\bf 2}) Second, slope-limiters and non-zero $\psi$ will add terms to $\divB$ which are not accounted for. 

Issue {\bf (1)} could be eliminated by solving {\em simultaneously} for all $(\dB)_{i}$. This is technically easy, but expensive (it amounts to a global sparse-matrix inversion every timestep), and ``overkill,'' because issue {\bf (2)} would still prevent $\divBi=0$. A much simpler, and more computationally efficient solution is to iteratively calculate $(\dB)_{i}$: 
\begin{align}
(\dB)&_{i}^{(n)} = (\dB)_{i,\,0}  -{\bf Q}\,\left(\frac{S_{0}^{(n)} + (\dB)_{i,\,0}:{\bf Q}}{{\bf Q}:{\bf Q}} \right) \\ 
S_{0}^{(n)} &\equiv - \sum_{j} \left[ {\bf B}_{i} + {\bf B}_{j} + (\dB)_{j}^{(n-1)} \cdot ({\bf x}_{ij} - {\bf x}_{j}) \right] \cdot {\bf A}_{ij}
\end{align}
where in the first-pass ($n=1$), we take $(\dB)_{j}^{(0)}$ to be the value of $(\dB)_{j}$ from the most recent time/drift step, calculate the new $(\dB)_{i}^{(1)}$ for all elements, then use these values to calculate the updated $(\dB)_{i}^{(2)}$, etc.

In practice, we find that given the errors sourced by the slope-limiters, we converge to nearly best-case accuracy in just two iterations ($n=2$). And since all the quantities here can be calculated in the same pass that is used to calculate $(\dB)_{i,\,0}$, the entire iteration series only requires one additional element sweep, compared to the ``standard'' method.

\begin{figure}
\plotonesize{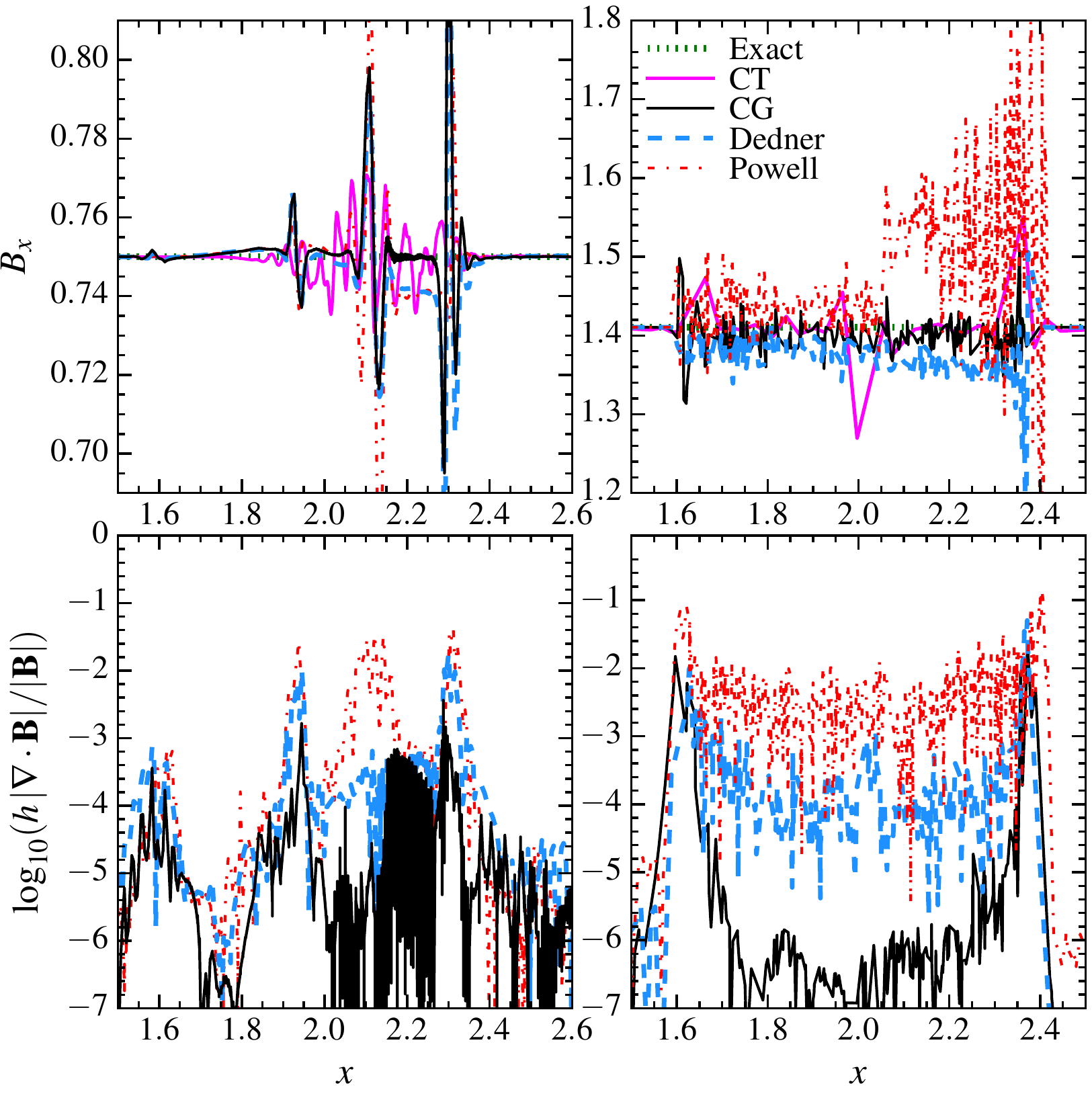}{0.99}
    \vspace{-0.25cm}
    \caption{Brio-Wu ({\em left}) and Toth ({\em right}) shocktubes (\S~\ref{sec:shocktubes}), at times $t=0.2$ and $t=0.08$, respectively. The setups for these and all other test problems follow \citet{hopkins:mhd.gizmo}. The tubes are 2D, with $\sim 256$ elements across the $\hat{x}$ (defined as the direction of shock propagation), with the initial grid mis-aligned from $\hat{x}$. We show the $\hat{x}$ component of ${\bf B}$ ({\em top}), which should be constant, and $\dberr$ ({\em bottom}), which measures the fractional magnitude of the magnetic divergence errors. All other fluid quantities ($P$, $u$, $\rho$, ${\bf v}$) tend to agree more closely between methods and are less sensitive to the divergence-control method. 
    With no divergence control, catastrophic errors overwhelm any solution (see Fig.~\ref{fig:shocktubes.bad}). Using only the \citet{powell:1999.8wave.cleaning} ``8-wave'' cleaning (``Powell''), $\dberr$ reaches $\sim0.1$, large noise/oscillations appear, and the incorrect shock jump is recovered, producing a systematic offset in $B_{x}$ (at $x\sim2.1-2.3$ and $x\sim2.0-2.4$, respectively). This offset does not decrease with resolution. Using the more sophisticated \citet{dedner:2002.divb.cleaning.scheme} cleaning reduces $\dberr$ by $\sim 1-2$\,dex, suppresses the oscillations, and dramatically reduces the systematic offset at the shock jump. However an offset still exists in both problems at the $\sim2-5\%$-level, which converges away slowly ($\propto N_{1D}^{-0.5}$). Our new constrained-gradient (CG) method maintains $\dberr < 0.01$ at the discontinuities, and $\dberr \ll 10^{-4}$ elsewhere; it produces still-smaller oscillations, and most important, completely eliminates the systematic offset at the jump. Constrained transport (CT) maintains $\dberr \lesssim 10^{-14}$ here; however, because the shock is not exactly grid-aligned, oscillations in $B_{x}$ still appear, comparable to CG.
        \vspacerpostplot 
    \label{fig:shocktubes}}
\end{figure}

\begin{figure}
\plotonesize{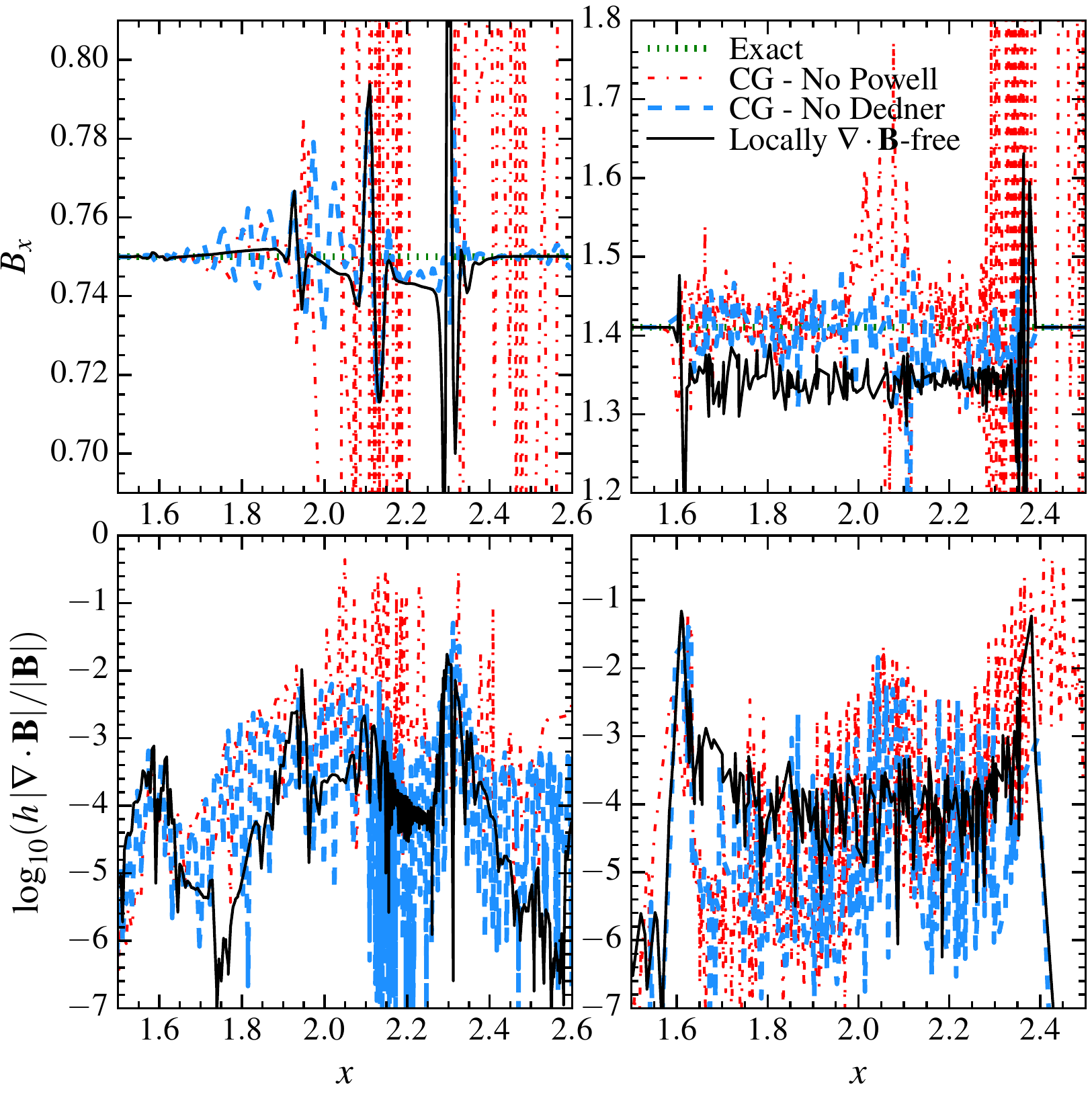}{0.99}
    \vspace{-0.25cm}
    \caption{Shocktubes from Fig.~\ref{fig:shocktubes}, with alternative divergence-control schemes. If, instead of approximating the global, divergence-free reconstruction according to Eq.~\ref{eqn:constraint}, we simply constrain the system to be ``locally divergence free'' (i.e.\ $\nabla\cdot{\bf B} = 0$ for the particle-centered gradient estimate; as \S~\ref{sec:local.divb.free}), we see there is essentially no reduction in the numerically problematic $\dberr$ term compared to the \citet{dedner:2002.divb.cleaning.scheme} without this constraint, and the systematic shock jump errors are actually increased. If we apply our CG scheme without the \citet{dedner:2002.divb.cleaning.scheme} cleaning, we recover the mean solutions but see large oscillations since the terms driving corrections to the gradients are not being damped. If we apply our CG scheme but ignore the \citet{powell:1999.8wave.cleaning} correction terms, the tensile instability appears and the oscillations grow to unacceptable levels.
        \vspacerpostplot 
    \label{fig:shocktubes.bad}}
\end{figure}

\begin{figure}
\plotonesize{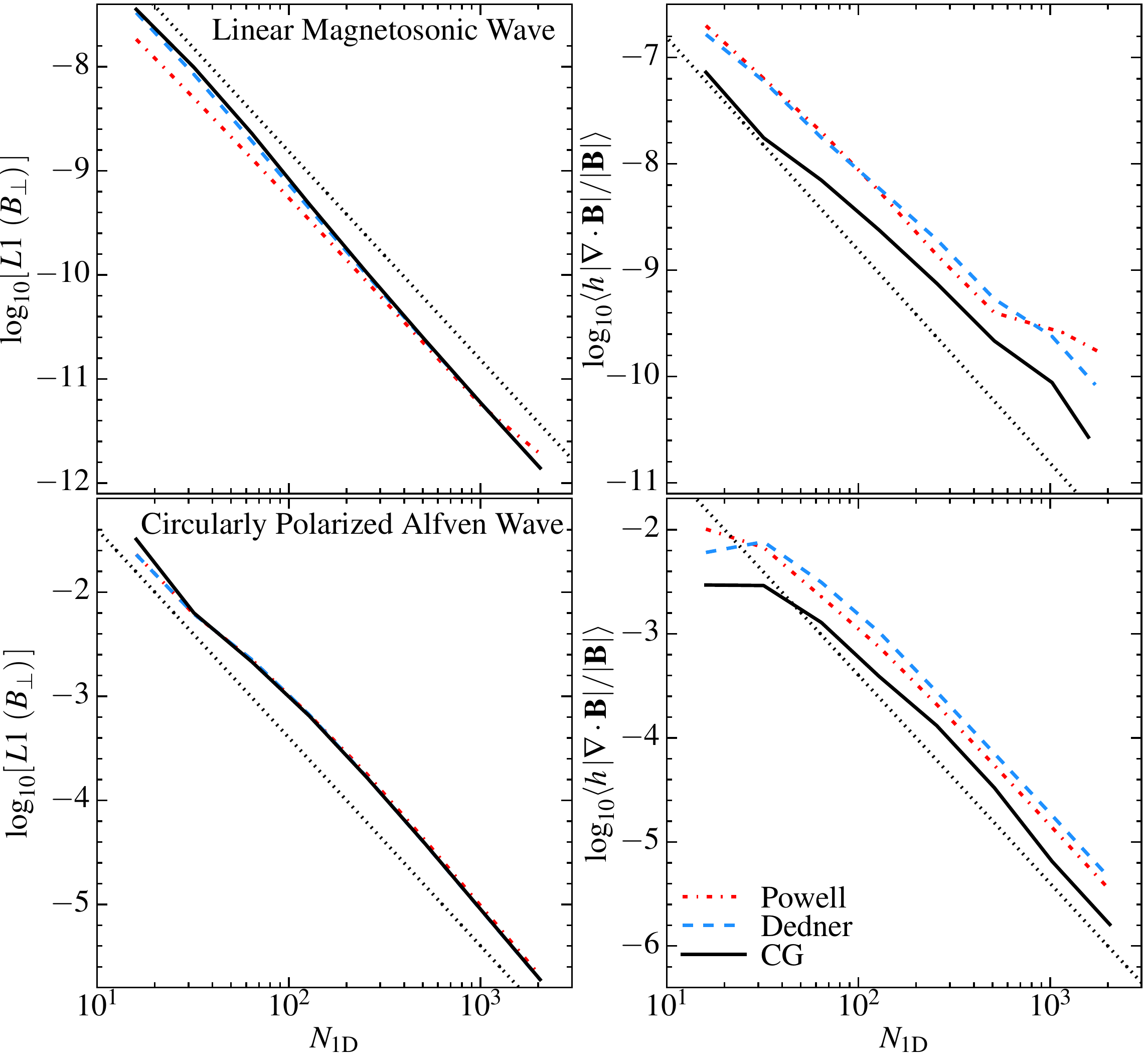}{0.99}
    \vspace{-0.25cm}
    \caption{Convergence study in MHD wave tests (\S~\ref{sec:waves}). 
    {\em Top:} Linear magnetosonic wave: a linear 1D fast magnetosonic wave is propagated one wavelength; we then plot the $L1$ norm ({\em left}; mean absolute error relative to the analytic solution) in magnetic field ($B_{\bot}$, direction perpendicular to propagation), and mean absolute divergence error $\langle \dberr \rangle$ ({\em right}; this is dominated by the largest errors in the domain, so the scaling is similar for ${\rm MAX}(\dberr)$).
    {\em Bottom:} Circularly polarized Alfven wave test: here the wave is an exact {\em non-linear} solution of the MHD equations. The wave is evolved in 3D and is tilted by $\sim 27^{\circ}$ relative to the $x$-axis, and errors are measured after is propagates five wavelengths. 
    All errors are plotted as a function of the number of elements across the domain $N_{1D}$; dotted lines show second-order convergence ($L1\propto N^{-2}$). We compare our CG, Dedner, and Powell methods; note that the CT results are from a different code with different convergence properties so a comparison here is not appropriate.
    In all cases, the methods here show roughly second-order convergence (slightly faster/slower for $B_{\bot}$ and $\dBerr$ in the linear wave test, but this is in the limit where the errors in some quantities approach floating-point accuracy).  As expected, the CG method systematically reduces $\dBerr$ (even moreso if we consider the median $\dBerr$). At low resolution, this comes at a small cost in accuracy (slightly larger errors in ${\bf B}$), owing to the constrained reconstruction of the magnetic field. However this appears to converge away quickly.
        \vspacerpostplot 
    \label{fig:waves}}
\end{figure}

\vspace{-0.5cm}
\subsection{Complications: Dealing with Slope Limiters and Cleaning Terms}

Issue {\bf (2)} is more challenging. The $\psi$ terms in Eq.~\ref{eqn:divb.definition} should be minimized along with $\divB$, so they are not particularly problematic. We can include them explicitly in our constraint solution for ${\bf G}$, using the same iterative approach to account for the fact that $\psi_{L}$, $\psi_{R}$, and $c_{h,\,ij}$ themselves depend on the gradients in the problem. However, this ``mixing'' of $\psi$ and ${\bf B}$ essentially defeats the purpose of the damping $\psi$ terms, and can introduce more serious numerical instabilities in the (rare) cases where $\psi/c_{h} \gg |{\bf B}|$. We therefore leave them, since their purpose is to damp $\divB$ where present. But we do reduce the contribution of the $\psi_{L}-\psi_{R}$ term in Eq.~\ref{eqn:divb.definition} by minimizing the least-squares deviation between $\psi$ extrapolated from the $i$ and $j$ ``sides'' {\em at the face locations}, rather than at the particle-$j$ locations (i.e.\ our preferred $\psi$ gradient is minimizes the squared deviation of $[\psi_{i} + (\nabla\psi)_{i}\cdot({\bf x}_{ij}-{\bf x}_{i})] - [\psi_{j} + (\nabla\psi)_{j}\cdot({\bf x}_{ij}-{\bf x}_{j})]$, rather than $\psi_{i} + (\nabla\psi)_{i}\cdot({\bf x}_{j}-{\bf x}_{i}) - \psi_{j}$). This is numerically consistent at the same order and trivial to implement using the same matrix based least-squares formulation, and we find it slightly reduces the divergences and improves the cleaning accuracy, so we use it throughout.

To deal with the slope-limiters $\tilde{\phi}$, we take advantage of our iterative approach. Generally speaking, there are two types of slope-limiters in most of the methods of interest. 

First, slope-limiters that are applied to the gradient after its calculation loop (and apply to all subsequent operations): $\tilde{\phi}$ is ${\rm MIN}(1,\,\tilde{\phi}^{\prime})$ where $\tilde{\phi}^{\prime}$ is chosen such that the reconstruction value of a field does not exceed the maximum/minimum neighbor values by more than some tolerance \citep{balsara:2004.second.order.accurate.mhd.schemes}.\footnote{In {\small GIZMO}, this takes the form: 
\begin{align}
\tilde{\phi}_{i}^{\prime} &\equiv {\rm MIN}{\Bigl[}1,\,\beta_{i}\,{\rm MIN}{\Bigl(}\frac{U^{\rm max}_{ij\,{\rm ngb}}-U_{i}}{U^{\rm max}_{ij,\,{\rm mid}}-U_{i}},\ \frac{U_{i}-U^{{\rm min}}_{ij,\,{\rm ngb}}}{U_{i}-U^{\rm min}_{ij,\,{\rm mid}}}{\Bigr)}{\Bigr]}
\end{align}
where $U^{{\rm max}}_{ij,\,{\rm ngb}}$ and $U^{{\rm min}}_{ij,\,{\rm ngb}}$ are the maximum and minimum values of $U_{j}$ among all neighbors $j$ of the particle $i$, and $U^{\rm max}_{ij,\,{\rm mid}}$, $U^{\rm min}_{ij,\,{\rm mid}}$ are the maximum and minimum values (over all pairs $ij$ of the $j$ neighbors of $i$) of $U$ re-constructed on the ``$i$ side'' of the interface between particles $i$ and $j$. The constant $\beta=1-2$ depending on local particle order. In our iterative CG implementation, we first limit with the ``normal'' $\beta=\beta_{0}$, then correct the gradient, then re-limit only if it exceeds the slightly weaker limiter with $\beta=2\,\beta_{0}$.} These are straightforward: we calculate our ``preferred'' gradient and then apply this limiter, and treat this as the new ``preferred'' gradient. After correction, the new gradient may violate this condition, so we can (optionally) re-limit it, and treat this as the new ``preferred'' gradient, and iterate until convergence (this iteration is outside the neighbor loop so has negligible cost). This converges to the gradient satisfying the desired slope limiter which comes as close as possible to the desired CG-corrected gradient. 

Second, another class of slope-limiters can (optionally) be additionally applied in pair-wise fashion between every interacting element pair in the flux computations; this ensures no local maxima/minima are created. Here the limiter $\tilde{\phi}_{ij}$ is unique to the element pair. We account for these limiters explicitly in our calculation of $S_{0}$: using both the current values of $(\dB)_{i}$ and $(\dB)_{j}$, we apply the limiters between each pair, and thus obtain a more accurate guess for the correction to $(\dB)_{i}$. 

These approaches allow us to handle arbitrary slope limiters and still return some valid result. But it is easy to see that application of any slope limiter can, under some circumstances, dis-allow the corrected value of $(\dB)_{i}$ needed to actually ensure $\divBi=0$. This is why our procedure ceases to significantly improve after a couple iterations. And clearly, a stricter slope-limiter prevents $\divBi$ correction under a wider range of circumstances. Therefore, when we implement our CG method, we ``weaken'' our normal pair-wise slope limiter for ${\bf B}$, to allow more flexible CG correction. This is important because, as we will show, if we do this and do not include any divergence-damping terms, it (unsurprisingly) produces large oscillations. Note, though, that we still retain the standard slope-limiter applied after the gradient calculation ($\tilde{\phi}^{\prime}$), so the second slope limiter is somewhat redundant anyways, and we only alter the slope limiters for ${\bf B}$.

\vspace{-0.5cm}
\subsection{Implementation}

We implement the CG method above in the code {\small GIZMO} \citep{hopkins:gizmo}. {\small GIZMO} is a mesh-free, finite-volume Godunov code, built on the gravity solver and domain decomposition algorithms of {\small GADGET-3} \citep{springel:gadget}. In \citet{hopkins:gizmo,hopkins:mhd.gizmo} we consider extensive surveys of test problems in both hydrodynamics and MHD (using the Dedner scheme) with this code, and demonstrate accuracy and convergence in good agreement with well-studied regular-mesh finite-volume Godunov methods. Because {\small GIZMO} is a multi-method code, it allows us to compare directly the effects of different divergence-control methods with an otherwise entirely identical code. Since we are not comparing hydro solvers here but the divergence control method, in what follows, we run {\small GIZMO} always in its MFM (Meshless Finite-Mass) mode, but we note that we have run several problems in the MFV (Meshless Finite-Volume) mode, and find nearly identical results (as expected from the comparisons in the methods paper). We have also implemented a limited, 2D version of CG in the public moving-mesh code {\small FVMHD3D} \citep{gaburov:2012.public.moving.mesh.code};\footnote{A public version of FVMHD3D is available at \\ \url{https://github.com/egaburov/fvmhd3d}} as expected from our previous comparisons, this is very similar to the {\small GIZMO} MFV results. For reasons shown below, when we implement our CG method, we still retain the Powell and Dedner source terms in the MHD equations, to deal with imperfect minimization of $\divBi$.

We have made public a version of the {\small GIZMO} code with the CG implementation used in this paper;\footnote{This is available at \gizmourl} users interested in the details of our implementation (for example, the exact numerical form of the slope-limiters used, kernel weights used for the least-squares calculation, etc.) are encouraged to examine the source code.

\begin{figure*}
\plotsidesize{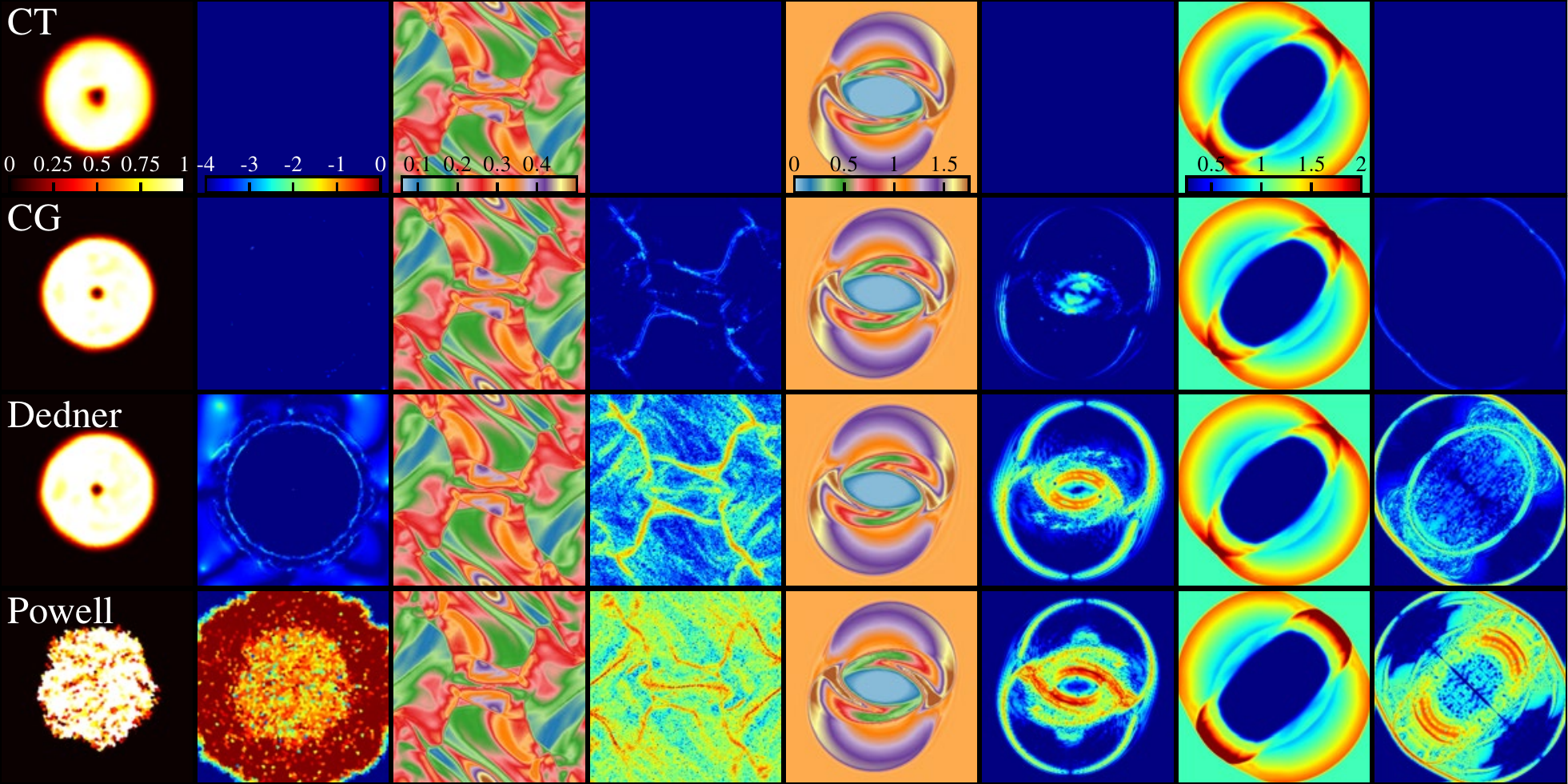}{1.02}
    \vspace{-0.25cm}
    \caption{Two-dimensional MHD tests. For each test (column), we compare four methods: CT, CG, Dedner, \&\ Powell (see Fig.~\ref{fig:shocktubes}), as labeled (top-to-bottom). Each pair of columns shows a map of a fluid quantity ({\em left}), and the corresponding map of $\log_{10}{(\dberr)}$ ({\em right}), with values following the colorbar. From left to right we show: {\em Left:} Field loop advection. We show magnetic pressure at time $t=20$. Methods should preserve a perfect circle at the maximum amplitude; numerical diffusion is visible at the center and edges. {\em Middle Left:} Orszag-Tang vortex, showing density at $t=0.5$. {\em Middle Right:} MHD rotor, showing gas pressure at $t=0.15$. {\em Right:} MHD blastwave, showing density at $t=0.2$. In each case Dedner, CG, and CT solutions are nearly-identical (the extra diffusion in CT in e.g.\ the field loop, is only because it uses an Eulerian, not Lagrangian code). The Powell scheme produces visibly incorrect features in the blastwave shock jump; and in the field loop test the $\dberr$ errors self-interfere and grow unstably. In all cases $\dberr$ decreases dramatically from Powell to Dedner and Dedner to CG, where it remains at values $\ll 0.01$. 
        \vspacerpostplot 
    \label{fig:cg.im.1}}
\end{figure*}

\begin{figure*}
\plotsidesize{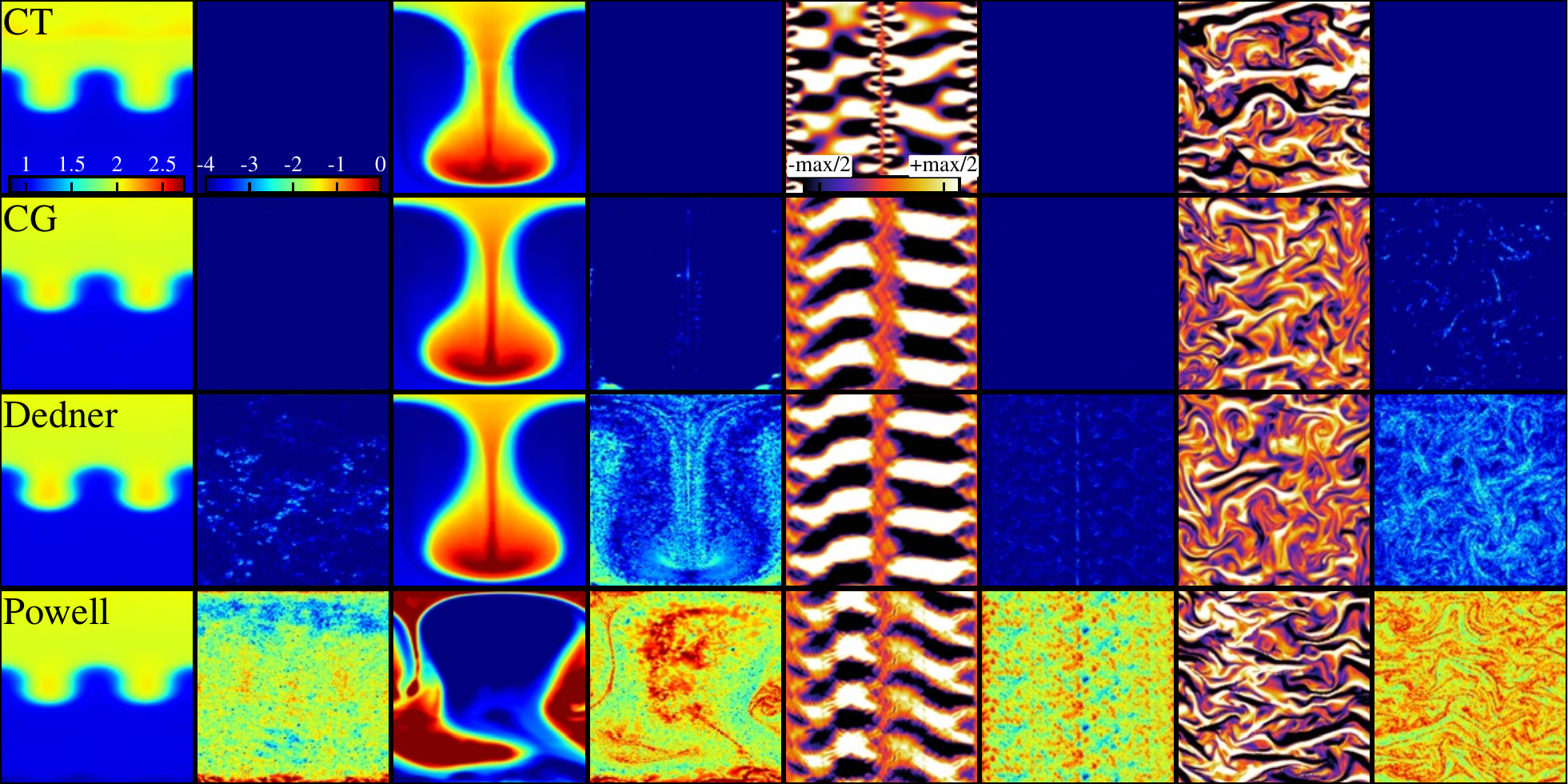}{1.02}
    \vspace{-0.25cm}
    \caption{Additional 2D tests, as Fig.~\ref{fig:cg.im.1}: MHD Rayleigh-Taylor instability (density is plotted, using a fixed scale given in the left-most colorbar) at times $t=6$ ({\em Left}) \&\ $t=16$ ({\em Middle Left}), and growth of the magnetorotational instability (MRI) in a shearing sheet at $t=10$ ({\em Middle Right}) and $t=19$ ({\em Right}). For the MRI we plot the azimuthal/toroidal component $B_{y}$ of the magnetic field, scaled relative to its maximum absolute value (``max''), so different times can be compared. All methods capture the linear growth of the RT and MRI, and breakup of the non-linear MRI into turbulence at late times. In each, Dedner, CG, and CT agree well (even into non-linear stages); CG maintains $\dberr\ll 0.01$ even well into non-linear and turbulent evolution. Powell schemes show small deviations in the MRI and linear RT growth, but the errors build up in the non-linear RT and destroy the solution. Note that the different pattern of modes in the CT MRI test owes to different implementations of shearing-box boundary conditions in the grid code used for the CT tests, as compared to the particle code for the other tests shown.
        \vspacerpostplot 
    \label{fig:cg.im.2}}
\end{figure*}

\begin{figure*}
\plotsidesize{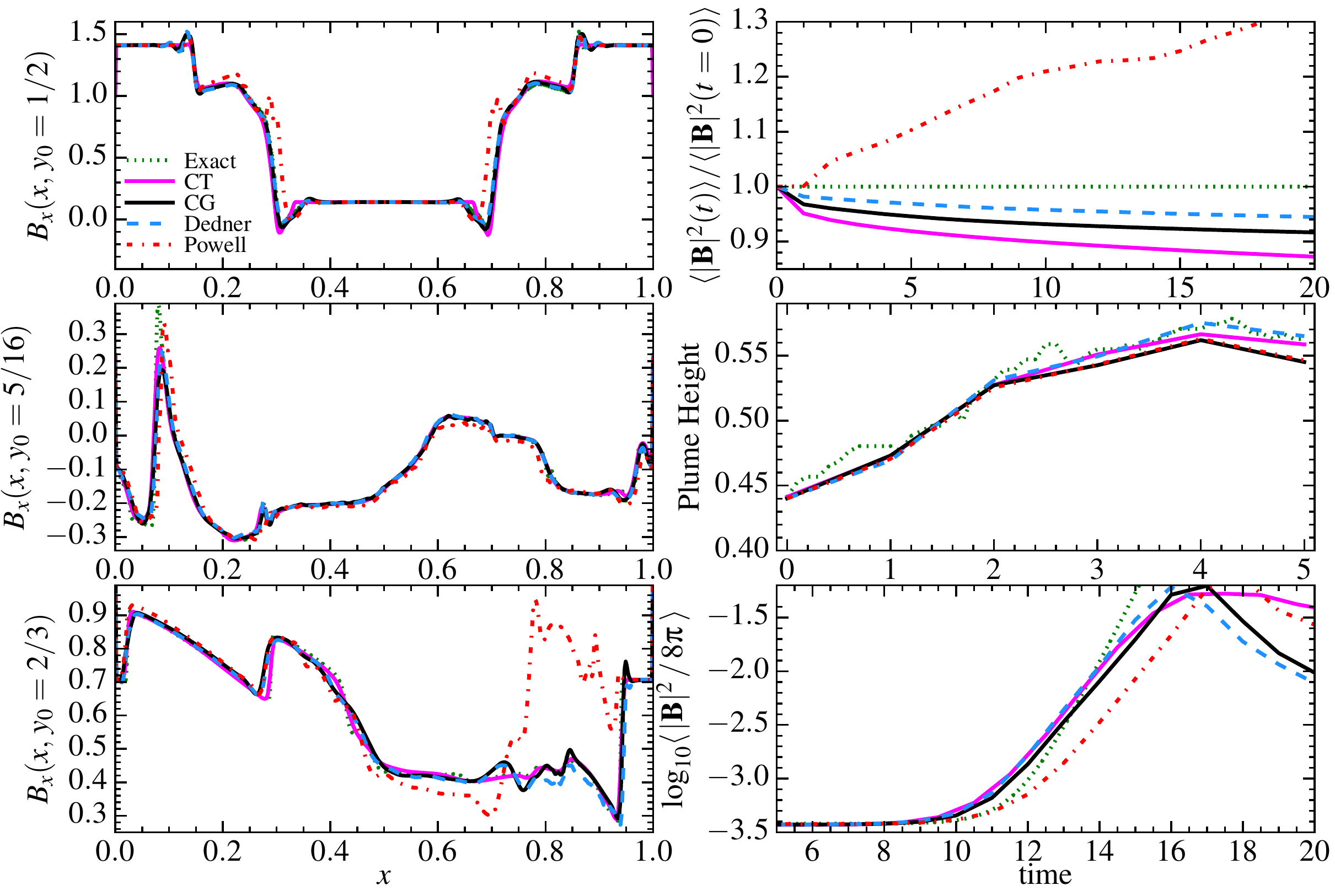}{0.95}
    \vspace{-0.25cm}
    \caption{Quantitative comparison of the 2D tests in Figs.~\ref{fig:cg.im.1}-\ref{fig:cg.im.2}. {\em Left:} Values of $B_{x}$ in horizontal slices, for the rotor ({\em top}), Orszag-Tang vortex ({\em middle}), and blastwave ({\em bottom}) tests (at the same time as Fig.~\ref{fig:cg.im.1}). We compare methods at $256^{2}$ resolution to an exact solution. All other fluid quantities show comparable or smaller deviations from the exact solution at this resolution.
    {\em Right:} Values versus time of the box-averaged $|{\bf B}|^{2}$ in the field loop test ({\em top}), low-density plume height in the RT test ({\em middle}), and magnetic energy density in the MRI test ({\em bottom}). In most tests Powell cleaning produces small deviations (offset shock positions, slower MRI growth); but in the blastwave and field loop tests the failure is dramatic. All other methods agree well and exhibit similar convergence rates. CG shows {\em slightly} smaller errors at fixed resolution compared to Dedner. Difference between CG \&\ CT (more diffusion for CT in the field loop \&\ RT tests, slightly sharper shock-capturing in CT in the blastwave, \&\ different late-time decay of the MRI) owe to the difference between grid methods (the CT results here) and Lagrangian methods (all others), {\em not} to divergence errors.
        \vspacerpostplot 
    \label{fig:cg.slices}}
\end{figure*}

\begin{figure}
    \plotonesize{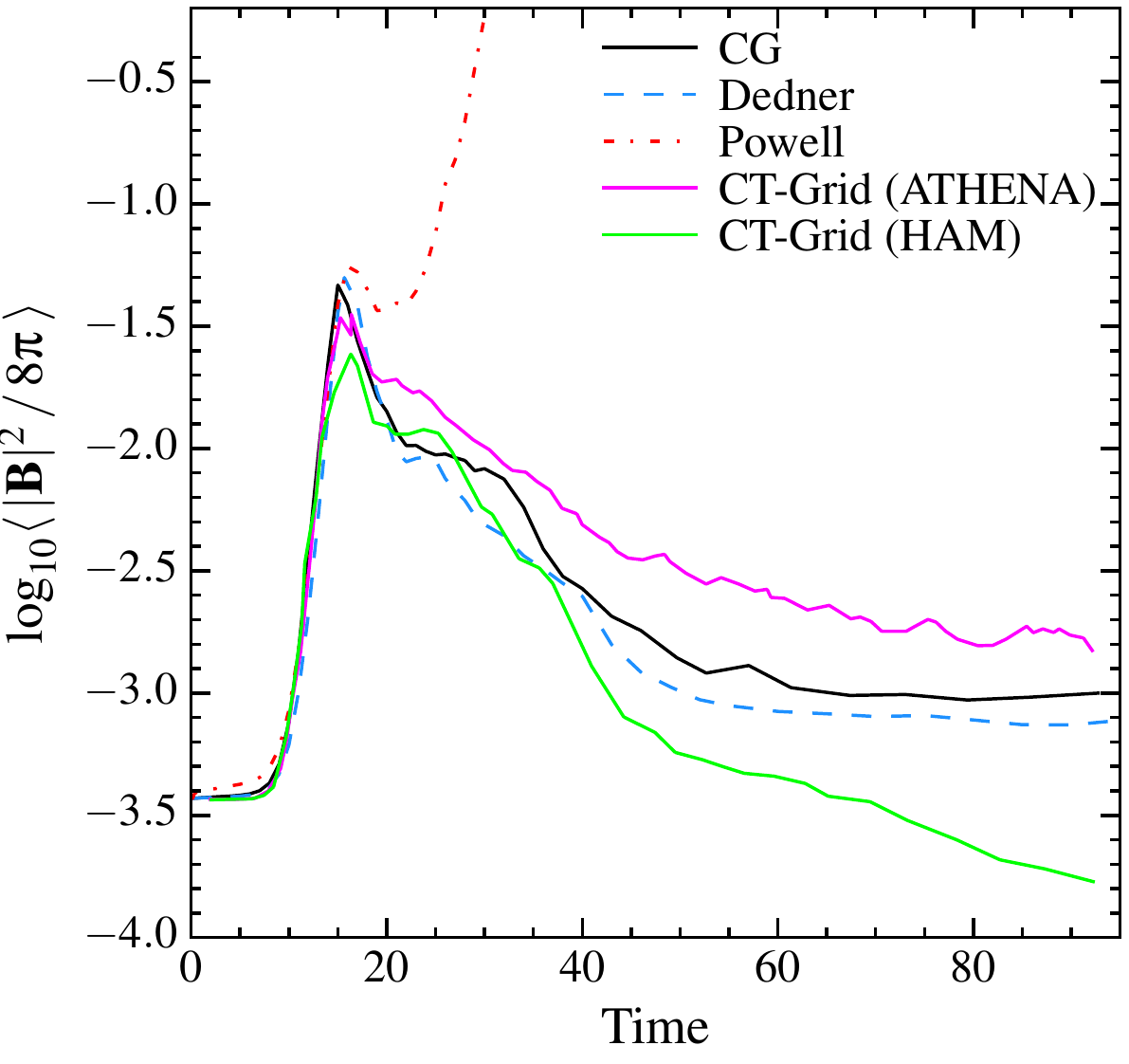}{0.9}
    \vspace{-0.25cm}
    \caption{Late-time evolution of the magnetic energy, for the MRI test in Figs.~\ref{fig:cg.im.2}-\ref{fig:cg.slices} (at $256^{2}$ resolution). Physically, the anti-dynamo theorem requires the magnetic energy decay; however the decay {\em rate} is known to be very sensitive to the numerical dissipation in a given code (greater dissipation producing faster decay). We compare our Powell, Dedner, and CG results in {\small GIZMO} to two different CT-based grid codes from \citet{guan:2008.shearing.box.mri} (the 2nd-order {\small HAM} code and the 3rd-order PPM unsplit CTU result from {\small ATHENA}). While the linear growth rates and peak amplitudes are similar, there are significant differences in the decay rate owing to differences in numerical dissipation. Our Dedner and CG results lie between the two CT grid results (with the higher-order CT result the least dissipative, as expected). Interestingly, CG is less dissipative than Dedner alone, even though it may produce a slightly less accurate reconstruction -- this owes to the reduction in $\dBerr$ producing less explicit dissipation from the Dedner divergence-transport and damping terms. The large errors in conservation from Powell-only cleaning lead to an unphysical non-linear runaway in $|{\bf B}|$ (in violation of the anti-dynamo theorem). 
    \vspacerpostplot 
    \label{fig:mri.late}}
\end{figure}

\begin{figure*}
 \begin{tabular}{cc}
  \includegraphics[width=0.87\columnwidth]{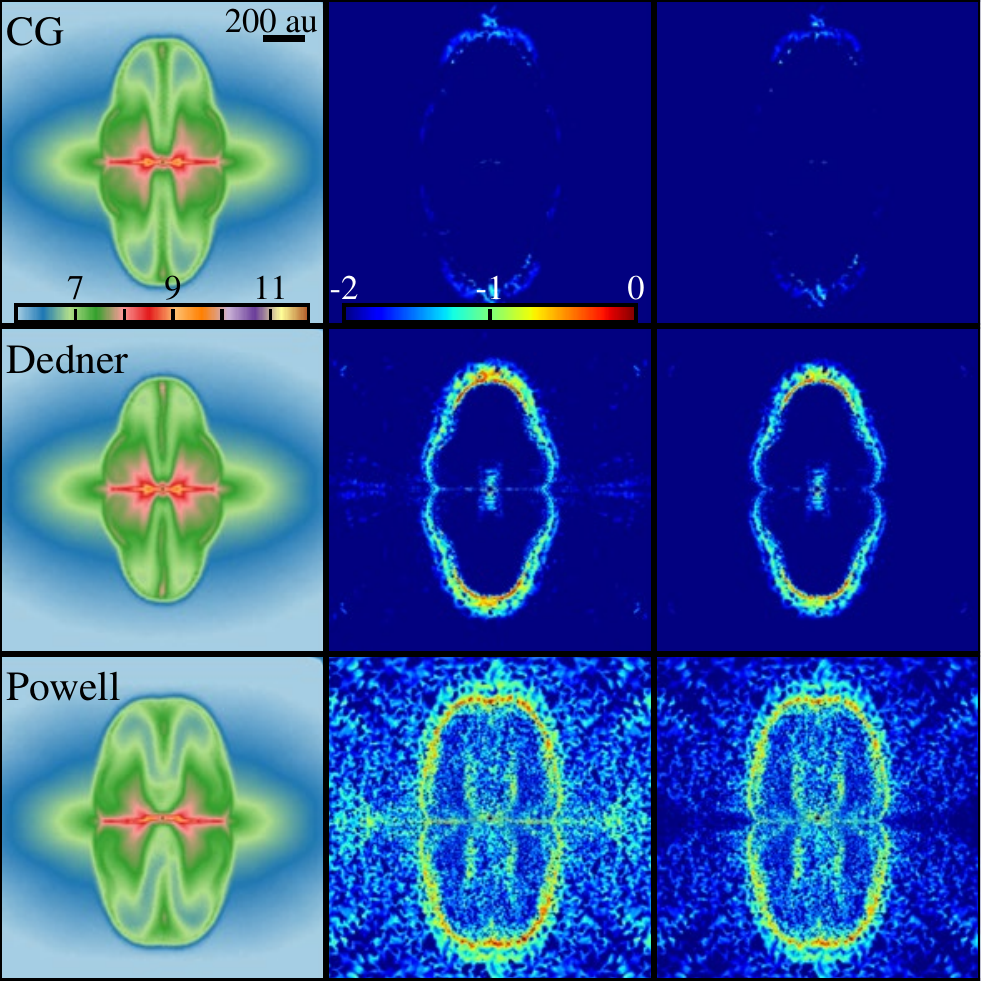} & 
  \includegraphics[width=0.87\columnwidth]{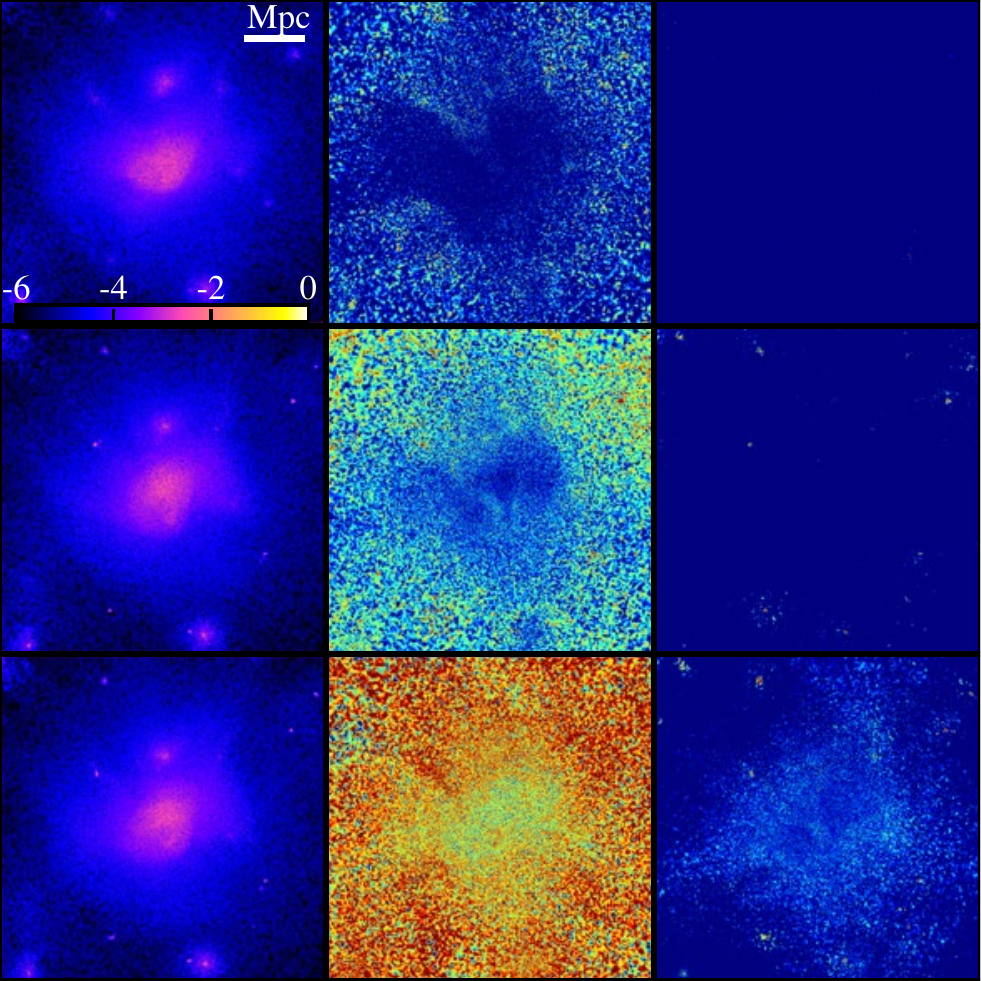} \\
  \includegraphics[width=0.87\columnwidth]{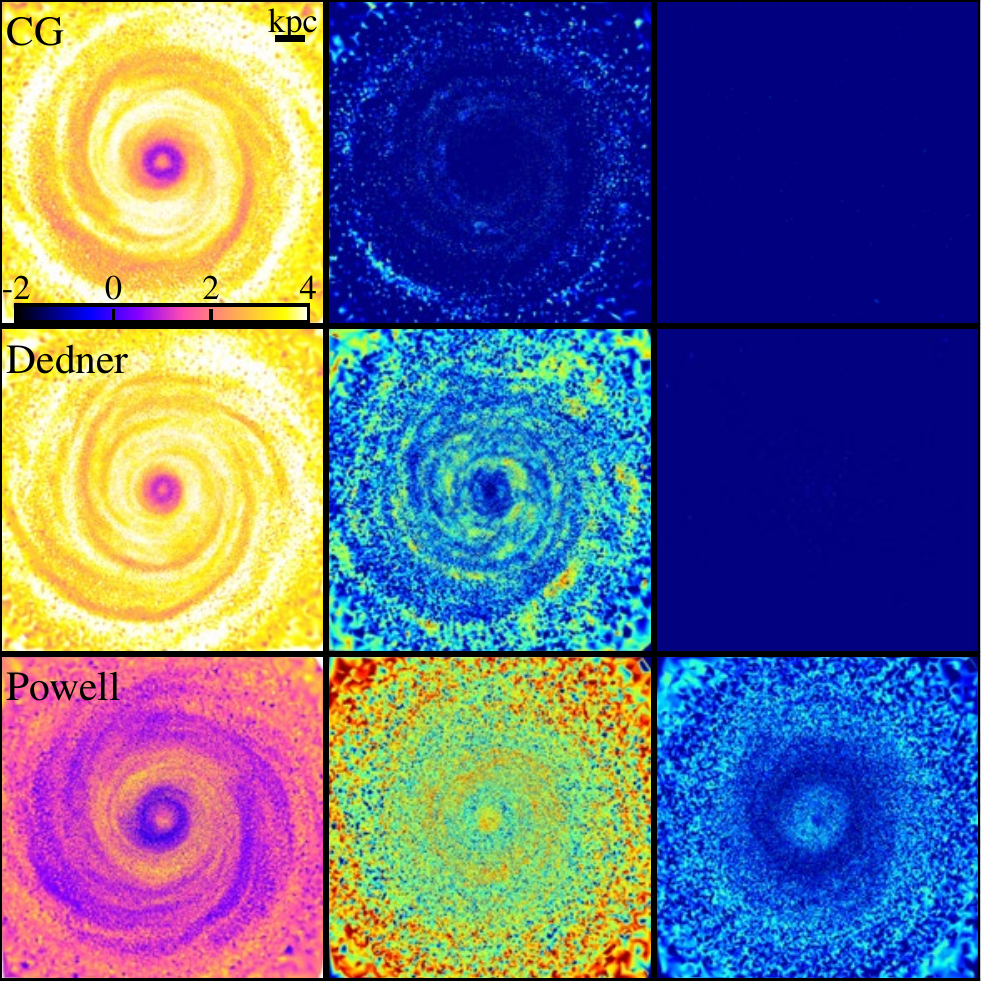} &  
  \includegraphics[width=0.87\columnwidth]{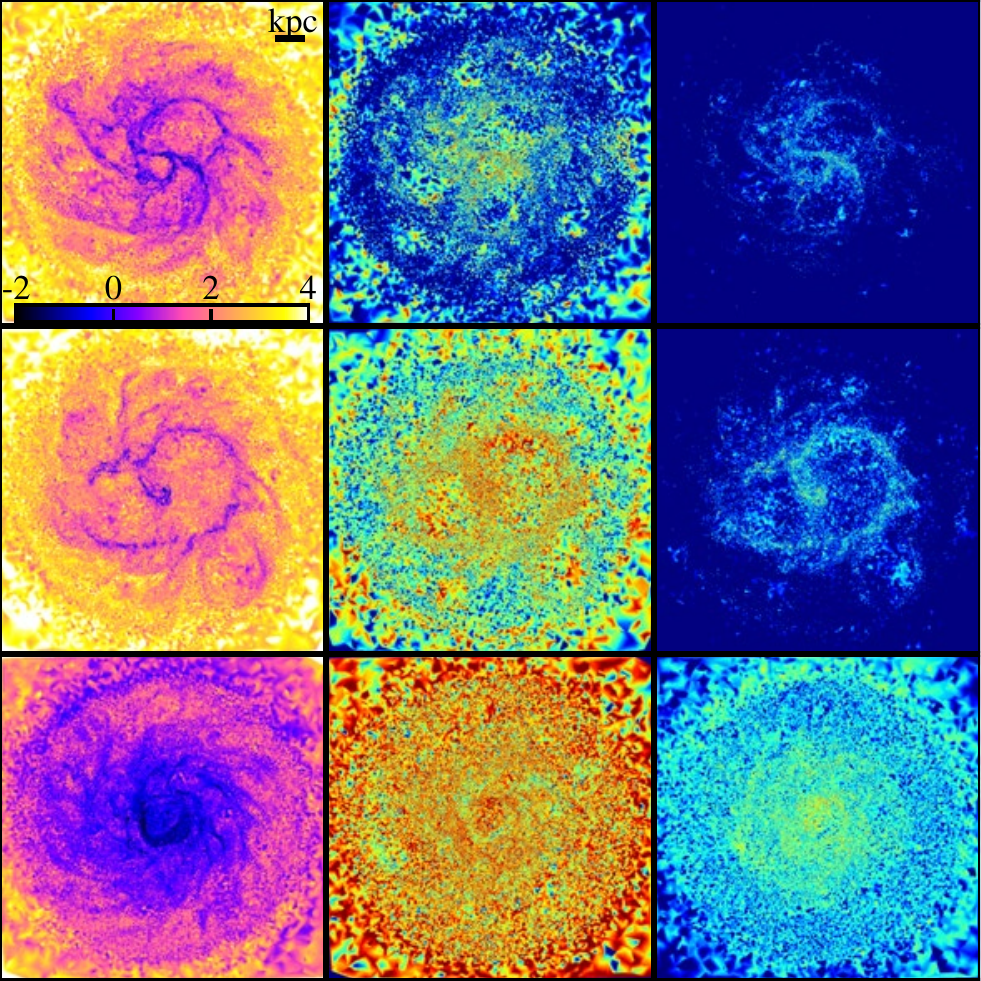}
 \end{tabular}
    \vspace{-0.25cm}
    \caption{3D, non-linear problems with self-gravity. For each we compare CG, Dedner, and Powell methods (rows, as labeled).
    {\em Top Left:} Jet formation via collapse of a rotating, magnetized protostellar core (after $\approx 1.1$ free-fall times). We plot the density ($\log_{10}[n/{\rm cm^{-3}}]$; {\em left}), divergence errors as before ($\dberr$; {\em middle}), and divergence error relative to the total gas pressure ($h_{i}|\nabla\cdot{\bf B}|_{i}/[{|{\bf B}|_{i}^{2} + 2\,P_{\rm thermal}}]^{1/2}$; {\em right}), in a slice through the jet axis. A protostar and rotating disk have formed, amplified ${\bf B}$, and launched a jet at this stage.     
    {\em Top Right:} Same quantities, for the magnetized Santa Barbara cluster (a cosmological, non-radiative simulation of dark matter and gas which forms a massive galaxy cluster-hosting halo), at redshift $z=0$ (slice through the cluster center shown). 
     {\em Bottom Left:} Isolated star-forming galaxy disk (gas, stars, and dark matter, slice through midplane shown) with radiative cooling and star formation, evolved for $500\,$Myr, with the smooth, sub-grid ``effective equation of state'' model for the ISM from \citet{springel:multiphase}. We plot plasma  $\beta \equiv P_{\rm magnetic}/P_{\rm thermal}$ ({\em left}), $\dberr$ ({\em middle}), and $h_{i}|\nabla\cdot{\bf B}|_{i}/[{|{\bf B}|_{i}^{2} + 2\,P_{\rm thermal}}]^{1/2}$ ({\em right}). 
    {\em Bottom Right:} Same quantities for the same disk, but evolved with the \citet{hopkins:2013.fire} FIRE models; these explicitly treat stellar radiation pressure \&\ photo-heating, stellar winds, and SNe, and resolve the multi-phase structure of the ISM and galactic winds. The multi-phase, turbulent structure and constant removal/addition of mass from the system makes this the most challenging case for divergence control. 
    The quantity $h_{i}|\nabla\cdot{\bf B}|_{i}/[{|{\bf B}|_{i}^{2} + 2\,P_{\rm thermal}}]^{1/2}$ demonstrates that the largest $\dberr$ values correspond to regions where ${\bf B}$ is dynamically irrelevant.
    In all cases, the simulations demonstrate the correct qualitative behaviors; the Dedner cleaning is acceptable, but CG further reduces $|\nabla\cdot {\bf B}|$ by $\gtrsim1$\,dex and maintains $h_{i}|\nabla\cdot{\bf B}|_{i}/[{|{\bf B}|_{i}^{2} + 2\,P_{\rm thermal}}]^{1/2} \lesssim 0.01$. However, the Powell cases are corrupted by large divergence errors: the jet is ``puffed out,'' has detached from the protostellar disk, and the protostar is migrating upwards out of the disk owing to momentum conservation errors associated with large $\dberr$. In the cluster and disk problems, the Powell-only $\dberr$ reaches $>1$, and ${\bf B}$-fields are amplified to order-of-magnitude too-large values.       \vspacerpostplot 
    \label{fig:cg.grav}}
\end{figure*}

\vspace{-0.5cm}
\section{Test Problems}
\label{sec:tests}

We now consider a series of test problems. Each of these has been studied in detail in \citet{hopkins:mhd.gizmo}, where we undertook a systematic comparison of different algorithms (MFM, MFV, SPH-MHD, and moving meshes, using the \citealt{dedner:2002.divb.cleaning.scheme} divergence-cleaning scheme, and grid/AMR schemes using constrained transport). Therefore we will not describe them in detail here.

\vspace{-0.5cm}
\subsection{Shocktubes}
\label{sec:shocktubes}

Fig.~\ref{fig:shocktubes} shows two shocktube standard shocktube tests: the sub-sonic, magnetically dominated \citet{brio:1988.shocktube} and super-sonic \citet{toth:2000.divB.constraint} shocktubes. We compare different divergence-control schemes, with resolution $\approx 256\times56$ across the domain plotted (a 2D grid, with the shock propagating at an angle $\pi/6$ to the grid).\footnote{Here and in all subsequent plots, the CT results are calculated with the grid code {\small ATHENA} \citep{stone:2008.athena}, run in its most accurate mode (PPM, CT, CTU). The accuracy and convergence properties of this code are well-studied. All other results are from {\small GIZMO}. Further comparison of the codes is found in \citet{hopkins:gizmo}.}

As discussed at length in \citet{hopkins:mhd.gizmo}, with no divergence-control at all, the schemes are unstable and crash (developing negative pressures). The minimal correction to restore stability is the \citet{powell:1999.8wave.cleaning} or ``8-wave'' cleaning; however this only subtracts the tensile terms from non-zero $\divB$, it does not actually control $\divB$; as a result we show therein that it produces incorrect shock jumps in $B_{x}$ and $u$, which lead to the shock being in the wrong position over time. Most importantly, these are zeroth-order errors which {\em do not converge} away at any resolution. This is known from previous studies as well \citep{toth:2000.divB.constraint,mignone:2010.ctu.mhd.divb.constraint,mocz:2014.constrained.transport.mhd}. 

Adding the \citet{dedner:2002.divb.cleaning.scheme} divergence-damping greatly reduces $\divB$ and allows the scheme to converge to the correct solution. With this scheme, almost every fluid quantity ($P$, $u$, $\rho$, ${\bf v}$, and $B_{y}$) has converged very well to the exact solution at this resolution. However, in $B_{x}$, which should be constant across $x$ in both shocktubes, we see some small, systematic offset from the exact solution still appear at this resolution. This owes to the relatively large $\divB$ which appears at the discontinuities. Unlike in the Powell scheme, this will converge away, but slowly (because the errors are low-order). 

Our CG scheme eliminates this systematic offset. We stress that every other fluid quantity is essentially indistinguishable from the result with the \citet{dedner:2002.divb.cleaning.scheme} scheme, in good agreement with the exact solution. There is still noise/oscillation associated with the discontinuity, but it returns to the correct systematic value. In fact, qualitatively similar noise appears even using a CT scheme; this owes to representing an inclined interface on a mesh (or non-aligned particle configuration), but this is less problematic because it converges away rapidly and does not lead to any systematic errors. 

We have also compared the \citet{ryu:1995.mhd.test.problems} shocktube, and 3D versions of the shocktubes; our conclusions are identical for the same types of discontinuities.

Fig.~\ref{fig:shocktubes.bad} considers some alternative formulations. First, instead of our CG method, we consider a ``locally divergence free'' projection as described in \S~\ref{sec:local.divb.free}. As we predicted, this does nothing to reduce $\divB$ or the systematic errors at the jumps (in fact they are worse), since it does not act on the problematic (non-local) terms. Next, we consider our CG method, but remove the \citet{dedner:2002.divb.cleaning.scheme} divergence-cleaning. While the average behavior is similar to the default CG case, and the systematic error at the shock disappears, we see very large oscillations in the post-shock solution. This is caused by the ``overshoots'' necessary at faces to obtain a divergence-free reconstruction sourcing additional $\divB$ waves, which cannot be damped. Finally, we consider the CG method but remove the cleaning and the \citet{powell:1999.8wave.cleaning} terms. Without these terms, even small $\divB$ errors can still grow unstably, and the system is specifically vulnerable to the tensile instability, which is triggered at the shock. As a result, the oscillations seen before grow unacceptably large. Therefore we will not consider these cases further.

\vspace{-0.5cm}
\subsection{MHD Waves}
\label{sec:waves}

While the shocktube problems above illustrate the differences between methods most dramatically, they are less useful as tests of accuracy and convergence (for which we desire smooth problems with known exact solutions). Fig.~\ref{fig:waves} shows a convergence study in two such problems. First, following \citet{stone:2008.athena}, we initialize a traveling fast magnetosonic wave with amplitude $\delta \rho / \rho = 10^{-6}$ (well in the linear regime) in a periodic domain of unit length (with background polytropic $\gamma=5/3$, density $\rho=1$, pressure $P=1/\gamma$, and magnetic field ${\bf B}/\sqrt{4\pi} = (1,\,\sqrt{2}, 1/2)$). After propagating one wavelength, the system should return to its original state so we define the error norm in the perpendicular field as $L1(B_{\bot}) = N^{-1}\, \sum_{i} |B_{\bot}(x_{i},\,t) - B_{\bot}(x_{i},\,t=0)|$. We similarly measure the absolute mean divergence errors. As shown in \paperone, the methods in {\small GIZMO} exhibit second-order convergence on this problem. Although our CG method does reduce the magnitude of $\dBerr$, the effect is only a factor $\sim 3$ -- all methods handle this problem accurately, owing to its relative simplicity and lack of discontinuities. More demanding is the circular polarized Alfven wave test from \citet{toth:2000.divB.constraint}; a 3D periodic box of unit size (particle number $N_{1D}^{3}$) with $\gamma=5/3$, $\rho=1$, $P=0.1$, $V_{\|}=0$, $B_{\|}=1$, $V_{\bot}=B_{\bot}=0.1\,\sin{(2\pi\,x_{\|})}$, $V_{z}=B_{z}=0.1\,\cos{(2\pi\,x_{\|})}$ (where $x_{\|}\equiv x\,\cos{\alpha}+y\,\sin{\alpha}$ and $\tan{\alpha}=1/2$ defines the angle of the wave propagation) is evolved until the wave propagates five wavelengths. We then measure the L1 norm in $B_{\bot}$ and mean $\dBerr$. This is inherently multi-dimensional and non-linear, making it more demanding. In all cases, there is some numerical wave damping, but again we see second-order convergence. Our CG method again reduces the mean $\dBerr$ by a similar factor. 

In both the linear and non-linear waves, we can see a slight increase in the error norms at low resolution with our CG method. In fact the Powell cleaning actually exhibits the smallest error norms on these problems. Provided the magnetic divergence errors are not so large as to corrupt the solution, this is the ``minimal correction'' so preserves the accuracy of the underlying method most faithfully. In the CG case, the (small) loss of accuracy owes to the constraint imposed in the reconstruction step -- so it cannot always reconstruct the ``most accurate'' (in a least-squares sense) gradients. This difference appears to converge away relatively quickly at higher resolution.

\vspace{-0.5cm}
\subsection{Dynamics Test Problems}
\label{sec:2d}

Figs.~\ref{fig:cg.im.1},\,\ref{fig:cg.im.2}, \&\ \ref{fig:cg.slices} show several 2D tests: advection of a field loop \citep{gardiner:2008.ctu.methods.paper}, the MHD rotor \citep{balsara:1999.mhd.rotor.test}, the \citet{orszag.tang.vortex} vortex, a strongly magnetized blastwave \citep{londrillo:mhd.blastwave}, the MHD Rayleigh-Taylor instability \citep[e.g.][]{jun.stone:rt.mhd}, and the development of the magneto-rotational instability (MRI) in a shearing-shear simulation \citep[following][]{guan:2008.shearing.box.mri}.  We show images of fluid quantities and the divergence errors, at some time into the non-linear evolution, and values of fluid quantities. All use $256^{2}$ resolution.

In each case, using the Powell scheme alone leads to {\em qualitatively} incorrect features (shocks in the wrong position, catastrophic noise, jumps with the wrong shape, etc), and $\divBnorm\sim1$; these are order-unity errors that do not converge away. In e.g.\ the RT instability, we see that while the linear (early time) behavior is reasonable, the non-linear behavior is destroyed by growing $\divBnorm$. In contrast, the Dedner-scheme, CG, and CT results are almost identical at this resolution, in the physical fluid quantities. However the Dedner scheme still produces some small regions where $\divBnorm$ can reach values as large as $\sim 0.01-0.1$, at this resolution (at sharp discontinuities). These deviations disappear in the CG result, which maintains {\em maximum} $\divBnorm \lesssim 0.01$, even at discontinuities, and mean $\langle \divBnorm \rangle$ about an order of magnitude smaller than the Dedner scheme. 

Quantitatively, Fig.~\ref{fig:cg.slices} illustrates the order-unity errors in shock positions and jumps that appear in the Powell scheme in the rotor and blastwave problems; it also illustrates in the field loop test that the scheme is numerically unstable, and the magnetic energy grows exponentially. The other methods agree well; the linear MRI and RT growth rate are almost identical at this resolution (and agree well with the analytic predictions). Note that in the field loop problem, the CG method is slightly more dissipative compared to Dedner (CT is more dissipative still, but this is primarily due to the CT code being an Eulerian, not Lagrangian, code). This is because there is a real non-zero $\divB$ set up in the ICs at the ``edge'' of the circle, as we numerically implement them. This forces the gradients to correct for this, and dissipate away the divergence.

In Fig.~\ref{fig:mri.late}, we expand our comparison of the magnetic energy in the MRI test to late times. The growth saturates and then the energy must decay according to the anti-dynamo theorem (in 2D, zero net flux simulations); but the decay rate is sensitive to the numerical diffusivity of the method \citep[see][and references therein]{guan:2008.shearing.box.mri}. We therefore compare our CG, Dedner and Powell results to two different CT implementations in different grid codes. We show this to emphasize that difference between the two CT codes is much larger than between our Dedner and CG results -- other factors (beyond the divergence-control scheme) dominate the numerical diffusivity (for a more detailed comparison of methods at different resolution, see \paperone). However, we do see slightly weaker numerical dissipation in our CG compared to (otherwise identical) our Dedner runs. This is surprising given the accuracy comparison in Fig.~\ref{fig:waves} (we might expect constrained gradients to be slightly less accurate and therefore more diffusive); however, the Dedner terms include explicit wave diffusion and damping sourced by $\nabla \cdot {\bf B}$, so the reduction in $\dBerr$ in CG leads (non-linearly) to a net reduction in numerical dissipation. More troublingly, in the Powell-only scheme, the large conservation errors that build up lead to a violation of the anti-dynamo theorem and unphysical late-time growth of $|{\bf B}|$.

Variations of the above tests with different seed ${\bf B}$-fields produce qualitatively identical conclusions. And because the {\small GIZMO} methods are Lagrangian, ``boosted'' or rotated versions of the tests are trivially identical to those shown. We have compared the MHD Kelvin-Helmholtz instability and ``blob'' test, but the qualitative differences between methods are identical to the RT test shown. The ``current sheet'' test in \citet{hawley:1995.mhd.tests} is a test of numerical stability which all methods here pass similarly (see \paperone); however as with the field loop we see greater dissipation in CG because the ICs contain a real non-zero $\divB$. We have also simulated low-resolution 3D versions of the RT, KH, field loop, and MRI problems with qualitatively similar results in all cases to the 2D tests here. 
 
As noted by \citet{gardiner:2008.ctu.methods.paper}, the field loop problem also provides a useful validation that we are minimizing the ``correct'' discrete representation of $\nabla \cdot {\bf B}$ for our scheme. The problem should at all times have $B_{z}={\bf B}\cdot \hat{z}=0$ (where the loop is in the $xy$-plane); if we initialize the problem with constant but non-zero $v_{z}$, then non-zero values of the discrete $\nabla \cdot {\bf B}\ne 0$ will produce growth in $B_{z}$. In the cleaning-based methods studied here, the \citet{powell:1999.8wave.cleaning} source terms should cancel these errors. Indeed, in each case (Powell, Dedner, and CG), we have run the problem with $v_{z}=1$ (comparable to the advection speed in the $xy$ plane) until $t=20$, and find $B_{z}\approx0$ to within machine errors ($\langle |B_{z}| \rangle <10^{-20}$) at all times after the first few timesteps. This verifies that the correct discrete source terms for the Powell cleaning are being applied (which are obtained using the {\em identical} definition of the discrete $\nabla \cdot {\bf B}$ used to derive the Dedner and CG terms, in turn). For further validation, we have re-run without the Powell source terms; with no cleaning, $\langle |B_{z}| \rangle$ grows exponentially (soon exceeding the initial field strengths). With the CG method (but no Powell term), $\langle |B_{z}|\rangle$ scales (as it should) $\sim h\,|\nabla \cdot {\bf B}|$, jumping to $\sim 10^{-7}$ in a few timesteps, but remaining around these values as long as we continue running.

\vspace{-0.5cm}
\subsection{Non-Linear Dynamics with External Forces}

Next we consider a set of non-linear dynamics problems where the dominant forces are not MHD, but gravity. These are especially challenging for divergence-control methods because elements are being constantly re-arranged by non-MHD forces in a manner that is often faster than the local fast magnetosonic crossing (hence fluid response) time. Unfortunately they are not rigorous tests, because exact solutions are not known, but they are useful validations that the code does not produce unphysical behaviors or new numerical instabilities under extreme conditions.

We consider (1) collapse of a rotating, self-gravitating proto-stellar core, to form a protostar and accretion disk which winds up a seed ${\bf B}$ field and launch an MHD jet, following \citet{hennebelle:2008.protostellar.jets}, (2) the MHD version of the ``Santa Barbara cluster,'' in which a cosmological simulation of gas+dark matter is followed using adiabatic (non-radiative) gas physics, from high redshift until the Lagrangian region being simulated forms an object with the mass of a galaxy cluster at $z=0$ \citep{frenk:1999.sb.cluster}, (3) an isolated (non-cosmological) galaxy disk, with gas, stars, and dark matter, radiative cooling, star formation, and stellar feedback, following the simple sub-grid \citet{springel:multiphase} ``effective equation of state'' model (in which the phase structure of the ISM is not resolved but replaced with a simple barytropic equation of state, as used in large-volume cosmological simulations; \citealt{vogelsberger:2013.illustris.model}), and (4) the same disk, treating feedback explicitly according to the FIRE (Feedback in Realistic Environments) project physics \citep{hopkins:rad.pressure.sf.fb,hopkins:stellar.fb.winds,hopkins:stellar.fb.mergers,hopkins:2013.fire,faucher-giguere:2014.fire.neutral.hydrogen.absorption}, which explicitly follow the multi-phase ISM, turbulence, and feedback from stellar winds, radiation, and supernovae. 

As noted above, these problems do not have known exact solutions, however there are specific qualitative behaviors that should be observed in each case (that must be present owing to basic considerations of conservation and/or dynamics). And they are valuable ``stress tests'' for divergence control (indeed, many algorithms cannot run these problems without crashing). Fig.~\ref{fig:cg.grav} summarizes the results. In all cases the behavior is very similar between our Dedner and CG results. In the jet test, we see the jet launched efficiently and the system evolves stably to late times; the mass-loading of these jets is in good agreement with much higher-resolution CT-based AMR results \citep[see][]{hennebelle:2008.protostellar.jets}, as discussed in detail in \citet{hopkins:mhd.gizmo}.\footnote{Unfortunately, we cannot rigorously compare CT methods in Fig.~\ref{fig:cg.grav}, since the relevant physics for these problems is not implemented in the same manner in any CT code. Moreover the gravity solvers are different for these codes, which (since gravity is the dominant force) can introduce much larger differences than the divergence-control method.} Likewise, in the Santa Barbara test, we see the cluster form, and non-linear field amplification in the cluster center; the profiles of density, temperature, magnetic field strength, and velocity also agree well with higher-resolution CT-based AMR runs \citep[compare][]{miniati:2011.mhd.charm.santabarbara}, up to subtle differences at small radii that depend on whether the methods used are Lagrangian or Eulerian (see \citealt{hopkins:gizmo}). In the disk problem, the disk remains smooth (as it should) with the ``effective EOS'' model, while with the FIRE model it rapidly develops super-sonic turbulence, multi-phase structure, and a strong galactic wind. In both cases the field is amplified; amplification is slower in the ``effective EOS'' case because (by construction) there is no sub-structure, turbulence, or galactic wind (so only global disk winding amplifies ${\bf B}$). In the FIRE case, the molecular gas tracing spiral structure and GMCs is clearly evident in the $\beta$ map as regions where the thermal pressure is sub-dominant. Although the qualitative behavior is similar, the CG method reduces $\dBerr$ by $\sim 1-2$\,dex, relative to the Dedner case (which, as expected, can reach large $\dBerr\gtrsim 0.1$ in sharp discontinuities). 

In all problems here, the Powell-only case is problematic: divergence errors reach $\dBerr \gg 1$, and even $h_{i}|\nabla\cdot{\bf B}|_{i}/[{|{\bf B}|_{i}^{2} + 2\,P_{\rm thermal}}]^{1/2} \sim 1$. This produces some qualitatively erroneous features: the jet disconnects from its launching zone, and the disk on small scales exhibits a ``bending'' because the protostar is actually moving with a net $z$-velocity that grows in time, until it ``self-ejects'' (owing to the momentum conservation error that comes with subtracting large $\dBerr$ terms); this is seen in SPH and grid-based codes using Powell-only control as well \citep[see][]{price:2012.sph.mhd.jets}. These should not be possible if the method preserves momentum and problem symmetry accurately. In the cluster and disk problems, the same errors seen in e.g.\ the field loop problem lead to artificial, unstable growth of magnetic energy: the ${\bf B}$ field is higher by about $\sim 1\,$dex at the times shown compared to the Dedner, CG, and CT solutions, and it grows at an unphysical rate (faster than any physical timescales such as the disk dynamical time, even in the smooth-disk case).

We have also compared the 3D MHD Zeldovich pancake test from \paperone\ with various initial ${\bf B}$ \citep[see][]{zeldovich:1970.pancakes,li:2008.cosmo.grid.mhd}. This test, unlike those above, actually has an exact solution and can be compared rigorously. Unfortunately, that is possible because of the highly simplified problem geometry which makes $\divB$ control unnecessary (even Powell schemes maintain $\divBnorm \ll 10^{-4}$), so (not surprisingly), all the methods here produce indistinguishable results from CT.

\vspace{-0.5cm}
\section{Discussion}
\label{sec:discussion}

We have introduced a new method, Constrained-Gradient MHD, to control the $\divB$ errors associated with numerical MHD. This involves an iterative approximation to the least-squares minimizing, globally divergence-free reconstruction of the fluid. We implement this in the code {\small GIZMO} and show that, compared to state-of-the-art MHD implementations using divergence-cleaning schemes, this is able to further reduce the $\divB$ errors by orders of magnitude, and improves numerical convergence and accuracy at fixed resolution. 

The performance cost of this method is small, compared to constrained transport on irregular grids. It requires one additional neighbor element sweep after the main gradient sweep (the iteration step), in which a the quantity $S_{0}$ is re-calculated based on the updated gradients. In our implementation in {\small GIZMO}, this increases the CPU cost of the hydro operations by $\sim 20-30\%$, with essentially no memory cost. 

This method is motivated in spirit by projection and locally divergence-free reconstruction schemes. However, it avoids the large overhead expense of projection schemes and does not modify the fluxes, and unlike traditional projection schemes it can operate with complicated slope limiters, non-linear gradient estimators, irregular mesh geometries, and adaptive timestepping, and preserves the convergence order of the code. Unlike locally divergence-free schemes, the method here actually minimizes the numerically unstable $\divB$ terms, as opposed to those from a different estimator.

We consider a large suite of test problems, and compare a variety of divergence-control schemes. With only Powell or ``8-wave'' cleaning, typical $\divBnorm\sim1$ in non-linear problems/discontinuities, and this sources zeroth-order systematic errors at discontinuities which do not decrease with increasing resolution. This causes serious problems on a wide range of problems: large systematic errors in shock jumps, catastrophic noise, exponential growth of magnetic energy, and order-unity violations of energy/momentum conservation and symmetry all appear even at high resolution. 

Adding a more sophisticated Dedner hyperbolic-parabolic cleaning-damping term allows the method to converge properly, and reduces $\divBnorm$ by $\sim1-2$ orders of magnitude. In most tests we find that this is sufficient for ideal behavior and high accuracy; however, non-negligible $\divB$ errors can still appear at large discontinuities, which lead to small systematic offsets in jump conditions that converge away relatively slowly. 

In every case, the CG method further reduces the problematic $\divB$ terms (by another $\sim1-3$ orders of magnitude). Critically, while the Dedner scheme requires some finite time and resolution to dissipate non-zero $\divB$, the CG scheme instantaneously controls $\divB$ across single resolution elements. As a result, in almost all cases, this means that the {\em maximum} $\divBnorm < 0.01$, even across arbitrarily sharp discontinuities. This eliminates the systematic offsets present in the Dedner scheme, and gives results which are, for all practical purposes in the tests here, indistinguishable from those obtained using CT methods which maintain $\divBnorm=0$ to machine precision. 

The CG method is easy to implement and quite general: it is applicable to any finite-volume or finite-element method with a well-defined reconstruction procedure (this includes our arbitrary Lagrangian-Eulerian MFM/MFV methods, Godunov and Galerkin-type finite-volume grid/AMR codes, moving-mesh methods, and finite-pointset methods). And it is independent of the ``default'' procedure used for gradient estimation: all that is needed is a sweep to calculate the terms needed for the correction tensor ${\bf G}$. Although the formal maintenance of $\divB=0$ is not as accurate as CT, it is compatible with arbitrary gradient definitions and slope limiters (these are usually implicitly limited to highly prescribed forms for the magnetic field in CT), is relatively inexpensive, and can trivially handle arbitrary mesh/point geometries, different numbers of spatial dimensions, and adaptive (non-uniform) timesteps. 

Finally, we have considered one simple implementation of the method, but it allows considerable freedom which merits further exploration. Different ``preferred'' gradients (we used a kernel-weighted least-squares estimate), penalty functions, iteration/convergence schemes (e.g.\ semi-implicit matrix methods), slope-limiters, and cleaning-field ($\psi$) terms in $S_{0}$ could be easily considered and may well provide superior performance. Higher-order generalizations should also be straightforward, although the equations in this paper would need to be modified for a higher-order reconstruction used in determining the $\divB$ error.

\vspace{-0.5cm}
\acknowledgments 
We thank our anonymous referee for a number of helpful suggestions and additional tests. Support for PFH was provided by the Gordon and Betty Moore Foundation through Grant \#776 to the Caltech Moore Center for Theoretical Cosmology and Physics, an Alfred P. Sloan Research Fellowship, NASA ATP Grant NNX14AH35G, and NSF Collaborative Research Grant \#1411920. Numerical calculations were run on the Caltech compute cluster ``Zwicky'' (NSF MRI award \#PHY-0960291) and allocation TG-AST130039 granted by the Extreme Science and Engineering Discovery Environment (XSEDE) supported by the NSF. 
\\

\vspace{-0.2cm}
\bibliography{/Users/phopkins/Dropbox/Public/ms}

\begin{appendix}
Stuff

\end{appendix}

\end{document}